\documentclass[twocolumn,journal]{IEEEtran}

\usepackage{amsfonts}
\usepackage{amsmath}
\usepackage{amsthm}
\usepackage{amssymb}
\usepackage{graphicx}
\usepackage{supertabular}
\usepackage{longtable}
\usepackage[usenames,dvipsnames]{color}
\usepackage{bbm}
\usepackage{fancyhdr}
\usepackage{breqn}
\usepackage{fixltx2e}
\usepackage{capt-of}
\usepackage{tikz}
\usetikzlibrary{matrix}
\usepackage{endnotes}
\usepackage{soul}
\usepackage{marginnote}

\newcommand{\mathsym}[1]{}
\newcommand{\unicode}[1]{}

\usepackage{float}

\usepackage{tabularx}

\usepackage[framemethod=TikZ]{mdframed}
\usepackage[framemethod=TikZ]{mdframed}
\usepackage{framed}
\mdfsetup{%
	outerlinewidth=1,skipabove=20pt,backgroundcolor=yellow!50, outerlinecolor=black,innertopmargin=0pt,splittopskip=\topskip,skipbelow=\baselineskip, skipabove=\baselineskip,ntheorem,roundcorner=5pt}

\mdtheorem[nobreak=true,outerlinewidth=1,%leftmargin=40,rightmargin=40,
backgroundcolor=yellow!50, outerlinecolor=black,innertopmargin=0pt,splittopskip=\topskip,skipbelow=\baselineskip, skipabove=\baselineskip,ntheorem,roundcorner=5pt,font=\itshape]{result}{Result}

\mdtheorem[nobreak=true,outerlinewidth=1,%leftmargin=40,rightmargin=40,
backgroundcolor=yellow!50, outerlinecolor=black,innertopmargin=0pt,splittopskip=\topskip,skipbelow=\baselineskip, skipabove=\baselineskip,ntheorem,roundcorner=5pt,font=\itshape]{theorem}{Theorem}

\mdtheorem[nobreak=true,outerlinewidth=1,%leftmargin=40,rightmargin=40,
backgroundcolor=gray!10, outerlinecolor=black,innertopmargin=0pt,splittopskip=\topskip,skipbelow=\baselineskip, skipabove=\baselineskip,ntheorem,roundcorner=5pt,font=\itshape]{remark}{Remark}

\mdtheorem[nobreak=true,outerlinewidth=1,%leftmargin=40,rightmargin=40,
backgroundcolor=pink!30, outerlinecolor=black,innertopmargin=0pt,splittopskip=\topskip,skipbelow=\baselineskip, skipabove=\baselineskip,ntheorem,roundcorner=5pt,font=\itshape]{quaestio}{Quaestio}

\mdtheorem[nobreak=true,outerlinewidth=1,%leftmargin=40,rightmargin=40,
backgroundcolor=yellow!50, outerlinecolor=black,innertopmargin=5pt,splittopskip=\topskip,skipbelow=\baselineskip, skipabove=\baselineskip,ntheorem,roundcorner=5pt,font=\itshape]{background}{Background}

\usepackage{hyperref}
\begin{document}
\title{{\color{Brown}
Working With Convex Responses: Antifragility From Finance to Oncology}}

\author{Nassim Nicholas Taleb\IEEEauthorrefmark{1}\IEEEauthorrefmark{3}and Jeffrey West\IEEEauthorrefmark{2}\\
    \IEEEauthorblockA{ 
   \IEEEauthorrefmark{1}Tandon School of Engineering, New York University\\
  \IEEEauthorrefmark{2}Integrated Mathematical Oncology, Moffitt Cancer Center\\
      \IEEEauthorrefmark{3}Corresponding author, nnt1@nyu.edu \\
 Forthcoming, \textit{Entropy}, 2023 }
}
\maketitle

% nd Jeffrey West $^{2}$\orcidA{}}
\begin{abstract}
	 We extend techniques and learnings about the stochastic properties of nonlinear responses from finance to medicine, particularly oncology where it can inform dosing and intervention. We define antifragility. We propose uses of risk analysis to medical problems, through the properties of nonlinear responses (convex or concave). We 1)  link the convexity/concavity of the dose-response function to the statistical properties of the results; 2)  define "antifragility" as a mathematical property for local beneficial convex responses and the generalization of "fragility" as its opposite, locally concave in the tails of the statistical distribution; 3) propose mathematically tractable relations between dosage, severity of conditions, and iatrogenics. In short we propose a framework to integrate the necessary consequences of nonlinearities in evidence-based oncology and more general clinical risk management.
\end{abstract}

%%%%%%%%%%%%%%%%%%%%%%%%%%%%%%%%%%%%%%%%%%

\section{ Introduction: Where the Idea of Antifragility Came From}\label{Background0}

%\subsection{What is fragility?}
%\subsection{Fragility heuristics in finance}
%
%\subsubsection{The survivorship bias argument}
%Why concavity is necessary under most classes of probability distributions
%
%\section{Medicine and Convexity}
%The "S curve" in dose response
%
%A review of second order effects as seen.
%
%\section{Oncology and Heuristics}
%
%Some practical heuristics
%
%\section{Medical risk management and therapeutic thresholds}
%
%\section{Conclusion}

The notions of fragility and antifragility were inspired from the payoffs and the intricacies of financial derivatives. The concept was introduced in Taleb (2012) \cite{taleb2012antifragile} and more formalized in Taleb and Douady (2013) \cite{taleb2013mathematical}.  While in the real world many phenomena are intuitively known to benefit from an increase in "volatility"\footnote{Usually and unless otherwise specified, the term \textit{volatility} maps to the standard deviation of the random variable, or the variability of a nonrandom one.}, only quantitative finance had names for such attributes, such as "long gamma"\footnote{The financial "derivative" contract has a positive local second mathematical derivative with respect to the underlying security.}, "long vega"\footnote{The financial derivative has a positive first derivative with respect to the standard deviation of the underlying security (or the returns of the security).}, and similar measures, always associated with some range of variation as these sensitivities are local and have themselves higher derivatives.\footnote{By "local" we refer to the fact that most payoff functions in finance are convex over a certain range, then linear or concave, with the second derivative changing in sign (the so-called "higher Greeks" in \cite{taleb1997dynamic} ).} Furthermore finance links some  nonlinear attributes of portfolios to the risk of "blowups" that is, a loss large enough to be irreversible, such as irrecoverable financial ruin. Such quantitative and qualitative models of ruin can give us a tractable generalizable definition of fragility. But centrally, Derivatives risk management, at its core, lies in distinguishing between the properties of a random variable $X$ and a payoff function $f(x)$, almost always nonlinear.

While Jensen's inequality (on which more in \ref{convex1}) is concerned with the first moment of the distribution, monotone convex (or concave) functions, and with a static expectation operator, financial payoffs are more complicated: the first static moment, while relevant, is not the sole focus as:
\begin{itemize}
\item The expectation must be conditioned on absence of "blowup", that is the left tail of the distribution must be constrained (see Geman et al, 2015)\cite{gemangeman}, which involves all higher moments of $f(x)$. 
\item The payoff functions are almost never monotone.
\item Taking into account higher moments of the distributions is analog to going beyond second order effects: third, fourth, etc.
\end{itemize}

\begin{figure} 
\begin{center}
\includegraphics[width=\linewidth]{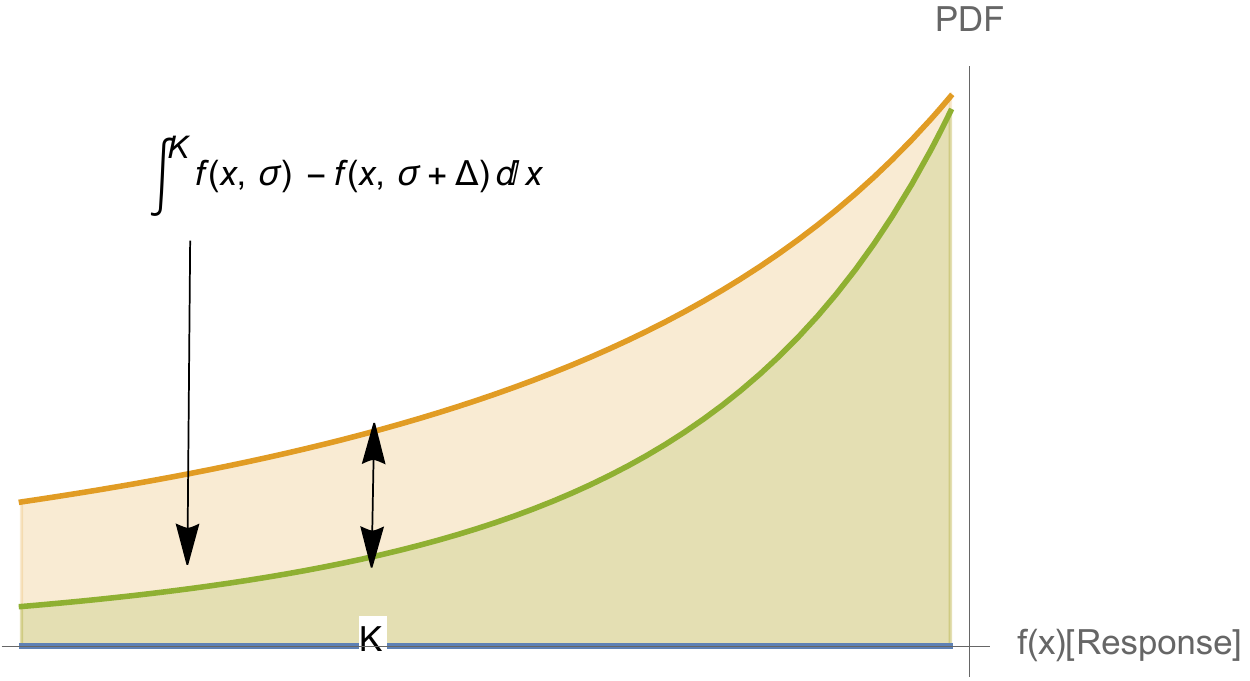}
\caption{Fragility below level $K$ as indicative of survival. It is not quite symmetric because global antifragility is conditioned on tail robustness ("to do well one must first survive"). The Taleb and Douady (2013)\cite{taleb2013mathematical} paper shows that the gap between $\int^K f(x, \sigma) dx$ and  $\int^K f(x, \sigma+\Delta) dx$ where $\sigma$ is the scale of the distribution is proportional to the concavity of $f(x)$. Hence without knowing the distribution (PDF above), one can gauge such effect by looking at the nonlinearity of $f(.)$ below the threshold $K$.}\label{tailfragility}
\end{center}
\end{figure}

\textbf{Fragility:} fragility as defined in Taleb (2012) and Taleb and Douady (2013)\cite{taleb2012antifragile},\cite{taleb2013mathematical}  is related to how a system suffers from the variability of its environment beyond a certain preset threshold (when the threshold is $K$, it is called $K$-fragility), see Figure \ref{tailfragility}, while antifragility refers to when it benefits from this variability —-in a similar way to as we saw what quantitative finance calls "vega" for an option or a nonlinear payoff, that is, its sensitivity to volatility or some similar measure of scale of a distribution. (Tail fragility maps to a risk of financial ruin, while local fragility does not necessarily mean ruin). In \cite{taleb2013mathematical}:
\begin{quotation}
	{\color{blue} \textit Simply, a coffee cup on a table suffers more from large deviations than from the cumulative effect of some shocks—conditional on being unbroken, it has to suffer more from "tail" events than regular ones around the center of the distribution, the ‘at-the-money’ category. This is the case of elements of nature that have survived: conditional on being in existence, then the class of events around the mean should matter considerably less than tail events, particularly when the probabilities decline faster than the inverse of the harm, which is the case of all used monomodal probability distributions. Further, what has exposure to tail events suffers from uncertainty; typically, when systems—a building, a bridge, a nuclear plant, an airplane, or a bank balance sheet—are made robust to a certain level of variability and stress but may fail or collapse if this level is exceeded, then they are particularly fragile to uncertainty about the distribution of the stressor, hence to model error, as this uncertainty increases the probability of dipping below the robustness level, bringing a higher probability of collapse. In the opposite case, the natural selection of an evolutionary process is particularly antifragile, indeed a more volatile environment increases the relative survival rate of robust species and eliminates those whose superiority over other species is highly dependent on environmental parameters.}
\end{quotation}

The paper above produced theorems linking the second derivative of $f(x)$ in some ranges of variation to sensitivity to the scale of the distribution of $X$. Thus the same sensitivity to the scale of the distribution can also express sensitivity to a stressor (dose increase) in medicine or other fields in its effect on either tail. Thus one single measure will allow us to express with comfortable precision the exposure to the \textit{disorder cluster}: (i) uncertainty, (ii) variability, (iii) imperfect, incomplete knowledge, (iv) chance, (v) chaos, (vi) volatility, (vii) disorder, (viii) entropy, (ix) time, (x) the unknown, (xi) randomness, (xii) turmoil, (xiii) stressor, (xiv) error, (xv) dispersion of outcomes\footnote{A positive (negative) sensitivity to one means positive (negative) to all others.}.Finally --and critically --the paper showed that one does not need to have an exact probability distribution to get a useful idea of the exposure since these metrics are based on acceleration, which washes out up to one order of magnitude the precision errors\footnote{For multivariate situations, an additive approach is used without any loss of effectiveness.}.

\textbf{Asymmetry Fragile/Antifragile:} The opposite of globally fragile (with respect to a random variable $X$) is not naively "antifragile", but both convex with respect to that variable and have a left tail constraint. In probabilistic representation $f(x)$ must have a positively skewed distribution. Furthermore, as with the fragile, antifragility is limited to a specific range of variations, and with respect to a single random variable.

The rest of this article will present medicine and convexity, then apply the notions of fragility-antifragility at two level: efficient dosing in oncology and an examination of iatrogenics as linked to convexity\footnote{We use by convention the term "convexity" or "convexity effect" in the presence of consequential nonlinearity, which can be concave: if the harm function is defined as positive it shows as convex, if negative, it shows as concave. Finance uses the expression "convexity bias" for both concave and convex responses (and with possible additional designation "positive" or "negative" convexity).}. Finally we present in the appendix an overview of convex responses in medicine.

%Tail properties (IMF PAPER) \cite{taleb2018IMF}

\section{ Medicine and Convexity}\label{Background}

\begin{figure}[t]
	\includegraphics[width=.65\linewidth]{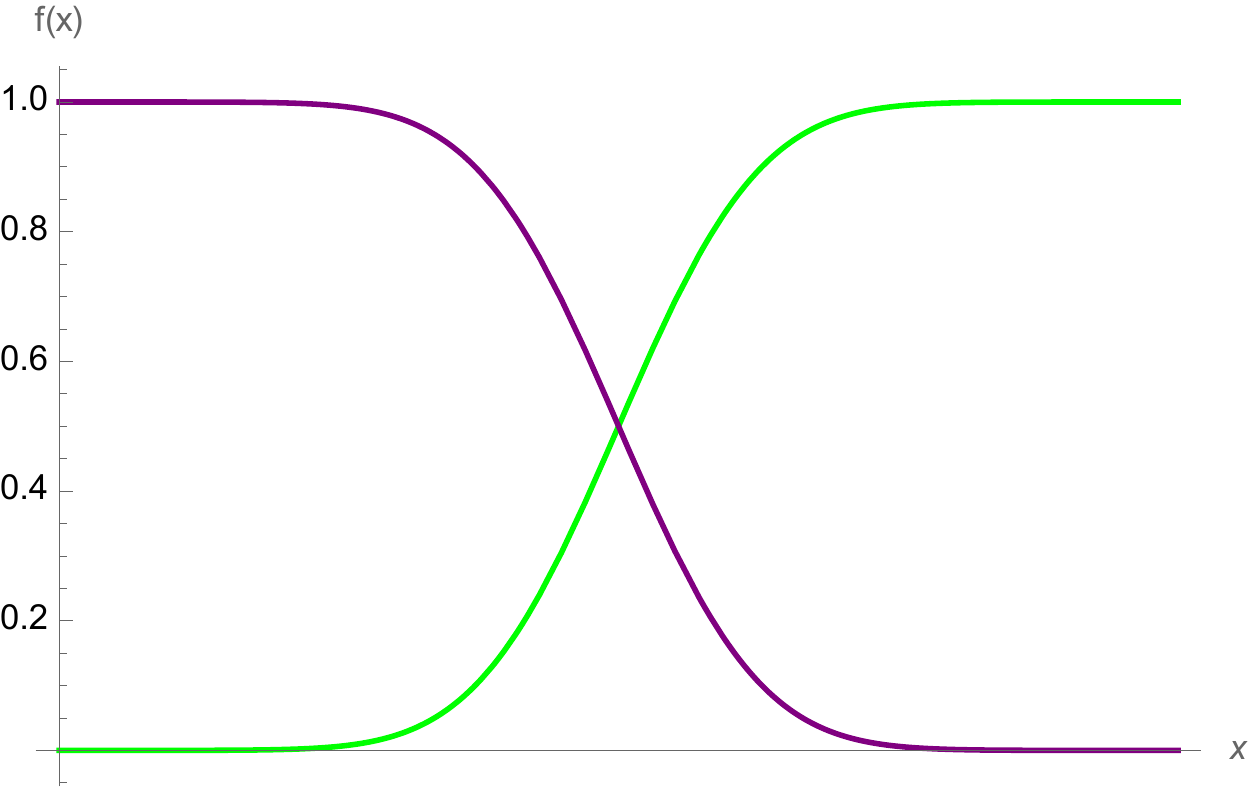}
	\caption{Simple (first order) nonincreasing or nondecreasing sigmoids, defined as floored and capped increasing functions. They map to the payoff in finance of a binary option with time left to expiration. As the sigmoid loses in smoothness (with the decrease time to expiration), it becomes at the limit a Heaviside function, see Fig.\ref{heaviside}.} \label{simplesigmoids}
	\begin{center}
	    \includegraphics[width=1\linewidth]{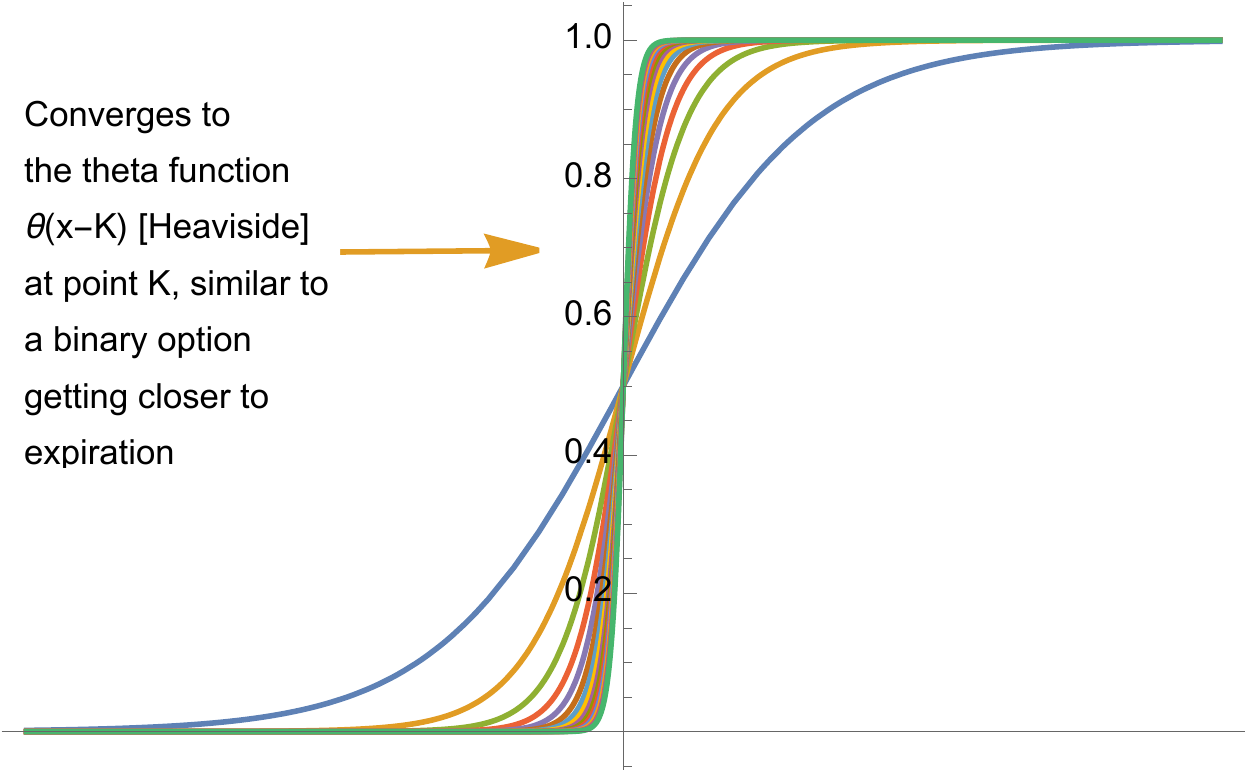}
    \caption{The smoothing of the Heaviside function as distribution or Schwartz function.} \label{heaviside}
	\end{center}
\end{figure}

Medicine has much simpler payoffs than quantitative finance. Most are generalizations around simple sigmoids, see Fig. \ref{simplesigmoids}, which were described in the mapping in Taleb and Douady (2013)\cite{taleb2013mathematical} as belonging to the benign class: the distribution of $f(x)$ is necessarily thinner tailed than that of $X$ owing to the boundedness of the function, dubbed more "binary" than "vanilla", see \cite{talebtechnicalincerto2020}.

However, in spite of such simplicity, little work in medicine has been done about probabilistic effect on convexity --almost always limited to first order effects and comparative statics. The probabilistic dimension of variability has been made explicitly in some medical domains, for instance there are a few studies connecting Jensen's inequality to patient responses with pulmonary ventilators: papers such as  Brewster et al. (2005)\cite{brewster2005convexity},  Amato et al. \cite{amato1998effect},  Funk (2004)\cite{funk2004comparison}, Arold et al. (2003)\cite{arold2003variable},  Graham et al.(2005) \cite{graham2005mathematical}, Mutch et al. (2007). To summarize the literature, continuous high pressures have been shown to be harmful (leading to increased mortality), but episodic spikes of ventilation pressures can be helpful with the recruitment of collapsed alveoli (natural breathing exhibits some variability, with some breaths deeper than others). However these papers stop at Jensen's inequality, and, further, explicit probabilistic formulations are still missing in other domains where the applications of these techniques are most needed, such as  intermittent fasting, episodic energy deficit, uneven distribution of sub-groups (say, proteins), vitamin absorption, moderate and low intensity training, fractional dosage, the comparative effects of low-intensity and distributed intervention vs intense and concentrated ones,  the chronic vs the acute, and similar effects\footnote{We note from the psychology literature the notion of overcompensation, see Den Hartigh and Hill (2022) \cite{den2022conceptualizing}.}.  The identification of convexity is still confined to local responses and did not generalize to decision-making under uncertainty and inferences concerning silent risks from the nonlinearity in dose-response. For instance the results did not reach the obvious relation between tumor size and the trade-offs of the intervention, or the extrapolation between the numbers needed to treat (NNT) and the potential severity of the side effects.
 
The connections we are investigating are necessary and mathematical: they work in both directions. We can illustrate as follows:
\begin{itemize}
\item a convex response to energy balance over a fixed time window necessarily implies gains from intermittent fasting in some situations and under some strict conditions (that is, higher variance in the distribution of nutrients) over some range within the limits of that time window,
\item the presence of metabolic problems in populations that have a steady supply of food intake, as well as evidence of human fitness to an environment that provides moderate variations in the availability of food, both necessarily imply a  concave response to food within a range and time frame.
\end{itemize}
%The point can be generalized in the same manner to energy deficits and the variance of the intensity of such deficits given a certain average. %\textit{Note the gist of our approach: we are not asserting that the benefits of intermittent dosage or the existence of a convex response are true; we are just showing that if one is true then the other one is necessarily so, and building decision-making policies that bridge the two.} 

Finally, a short summary of the above is as follows. Convexity analysis in medicine is at two dimensions, first, working with the nonlinearity of dosing, second, for risk analysis for patients and groups.

\textbf{Missing second order effect:} One frequent lacuna in the literature is ignoring the second order effects when making statements derived from from empirical data. One example is dietary recommendations for food group composition rather than frequency. Epidemiological interpretations of the Cretan diet relied solely on composition\footnote{A simple intuition of the second order effect in nutrition is as follows. Eating once a day vs., say, three presents a difference if the response function is nonlinear: an average of response functions is not a response function of an average. }. But frequency matters: the Eastern Orthodox Church has, with minor local variations, around two hundred days of vegan fasts per year. This is an episodic protein deprivation; fatty meats are consumed in lumps (on Sundays and holidays) which compensate for such a deprivation (recall the threshold in Fig.\ref{Jensen}. As shown in the literature review in Appendix A, there is a need for a mathematical bridge between studies of \textit{variability}, say Martin et al.(2006) \cite{martin2006caloric} and Fontana et al (2008)\cite{fontana2008long} on one hand, and the focus on \textit{composition} --the Longo and Fontana studies, furthermore, narrows the effect of the frequency to a given food type, namely proteins\footnote{Lee and Longo (2011) \cite{lee2011fasting} "In the prokaryote E. coli, lack of glucose or nitrogen (comparable to protein restriction in mammals) increase resistance to high levels of $H_2 O_2$ (15 mm) (Jenkins et al., 1988) \cite{Jenkins1988starvation}"}. Further, the computation of the "recommended daily" units may vary markedly if one assumes necessary stochasticity.

\textbf{Extracting past statistical attributes and frequencies:} A central question is that if we need a certain dose of stressors, whether in intermittence of nutrition or necessary exercise, these might represent the attributes of an "ideal" environment. Whether evolutionary or not, this is the one to whose stochastic properties we are most adapted. We can therefore reverse engineer the stochastic nature of such an "ideal" environment by finding the various conditions that result from reduction of stressors.  We noted that papers such as Kaiser (2003) \cite{kaiser2003sipping} and Calabrese and Baldwin (2003), \cite{calabrese2003hormesis} do not bridge the results to the point that hormesis may correspond to a "fitness dose", beyond and below which one departs from such ideal dispersion of the dose $x$ per time period. 

Such a reverse engineering uses the visible dose-response curve to make inferences about the ideal parametrization of the probability distribution of nutritional balance and vice-versa. For example, assessing the benefits of episodic fasting and the length of windows for neoplasms, insulin resistance, and other conditions can lead to understanding some kind of "fitness" to an environment endowed with a certain structure of randomness, either with the $\sigma$ above or some more sophisticated attributes of the probability distributions (such as higher moments hence different shapes). For example, if insulin resistance can be reduced thanks to occasional deprivation (a certain variance), say one 24 hour fast every week, 3 days of fasting per trimester, and a complete week every five years, then we can extract and parametrize a probability distribution of ancestral deprivations. A comprehension of the exact mechanism by which such intermittences works can be helpful but is not needed given the robustness of the mathematical connection between functional and probabilistic.

\begin{figure*}[t]
\includegraphics[width=1\linewidth]{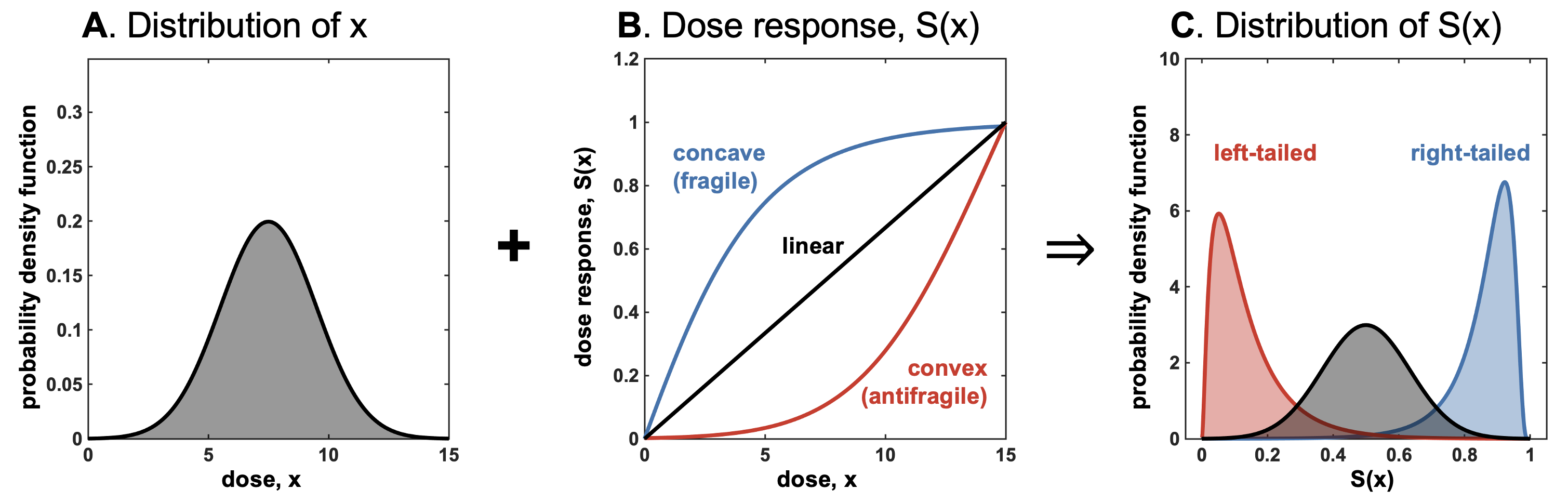}	
\caption{These three graphs (related to the convex (concave) transformations of random variables) summarize and simplify our main idea; they show how we can go from the reaction or dose-response $S(x)$, combined with the probability distribution of $x$, to the probability distribution of $S(x)$ and its properties: mean, expected benefits or harm, variance of $S(x)$. Thus we can play with the various parameters that can affect $S(x)$ and those that can affect the  distribution of $x$, and extract results from the output. $S(x)$ as we show can take different forms (We chose a monotone convex or concave $S(x)$ but can also use a second order mixed sigmoid). }\label{DoseResponseDist}
\end{figure*}

% REWRITE

\begin{figure*}[t]
    \includegraphics[width=\linewidth]{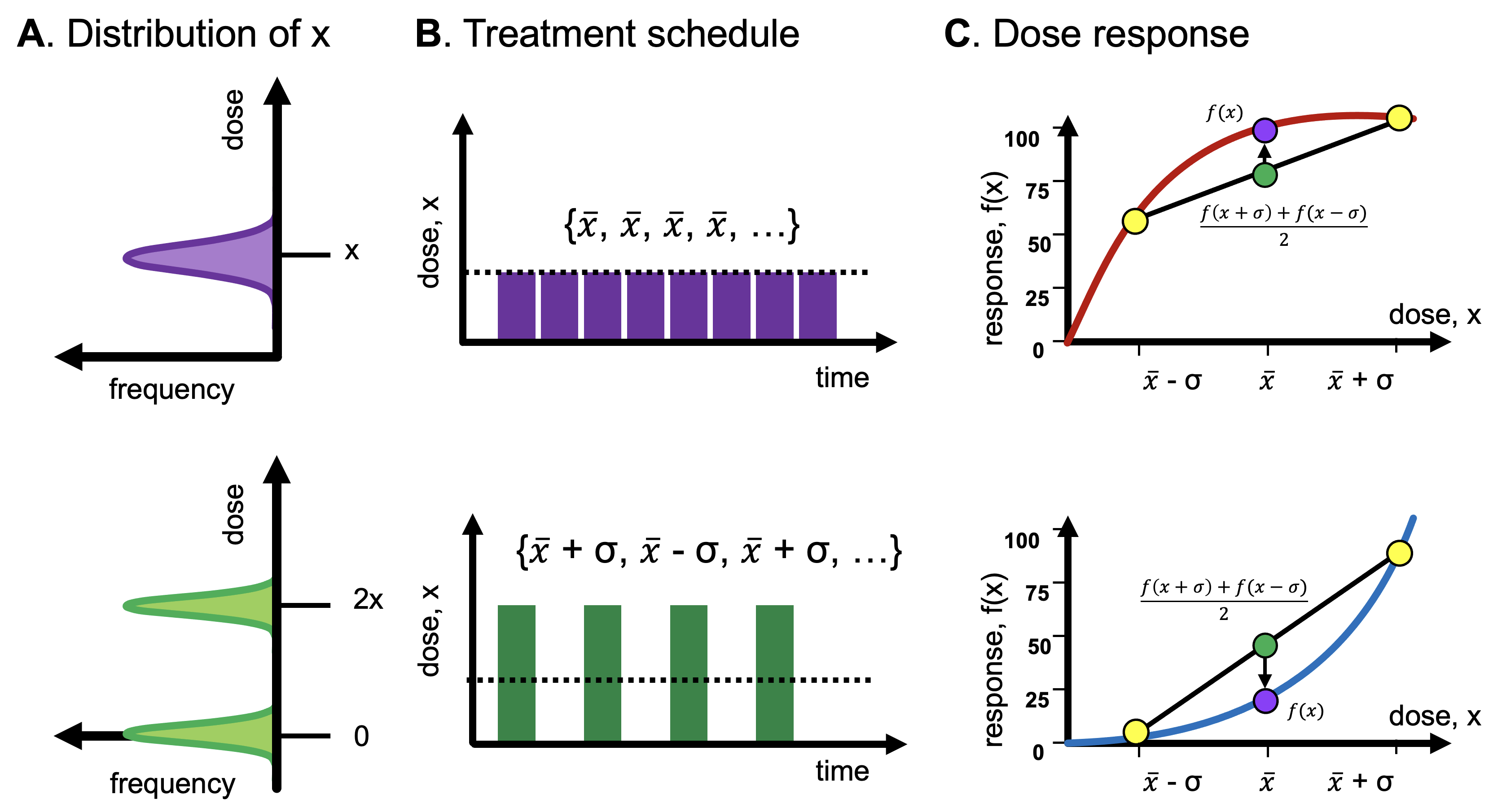}
	\caption{Example treatment scheduling protocols. (A) Input distribution of dosing is typically unimodal (``even'') or bimodal (``uneven''). (B) Protocols are typically fixed, with doses administered at regular intervals. It may be feasible to temporarily increase the dose (green), with periodic treatment holidays. (C) Even treatment is optimal to maximize response for concavity; uneven for convexity. } \label{Jensen}
\end{figure*}

\subsection{Antifragility in treatment scheduling} 
In figure \ref{DoseResponseDist}, the input distribution of dose $x$ is subject to the convexity (or concavity or linearity) of the dose response function, which influences the tail of the outcome distribution. Importantly, an oncologist has ``first-mover'' advantage \cite{stavnkova2019optimizing} and has the benefit of prescribing an ``even'' treatment protocol with no variance (figure \ref{Jensen}, top row) or an ``uneven'' treatment protocol with positive variance (figure \ref{Jensen}, bottom row). 

In medical practice, treatment protocols are typically fixed with doses administered at regular intervals (e.g. figure \ref{Jensen}). The distribution of dosing is unimodal (purple; continuous dosing), or at most bimodal (green; intermittent dosing). Manipulation of dose volatility when designing treatment protocols is under-utilized as a strategy in cancer treatment. In place of a dose $x$, one can give, say, 120\% of $x$, then 80\% of $x$, with a more favorable outcome one is in a zone that benefits from unevenness. If antifragile, more unevenness is more beneficial: 140\% followed by 60\% produces better effects \cite{west2021antifragile}.

\section{Antifragility in oncology}
Across all treatment modalities, cancer treatment is intended to induce perturbations to environmental conditions within a tumor leading to cell death, altering vasculature, or impacting immune response. However, the most common treatment paradigm is the ``maximum-tolerable dose'' (MTD) dosing protocol, whereby the dose is maximized, and only limited by tolerability, toxicity, and side-effects. To re-phrase, oncology research is implicitly focused maximizing ``first-order'' treatment effects by increasing the cumulative dose \cite{piccart2000impact} or shortening the time between dose\cite{citron2008dose}. The ``log-kill'' law proposes an MTD protocol for cytotoxic chemotherapy agents that decreases the amount of time over which a cumulative dose is delivered as toxicity allows\cite{skipper1964experimental}. More recently, metronomic therapy proposes frequent, low doses known to provide an anti-angiogenic effect during chemotherapy -- still implicitly optimizing a first-order effect of cumulative dose:  \cite{kerbel2004anti}.

Oncology must consider convexity in strategizing treatment protocols. Although convexity was not a consideration initial clinical design, recent approaches have had success in managing second-order effects through the practice of high / low dosing. Intermittent high dosing of tyrosine kinase inhibitors (TKI) in HER2-driven breast cancers was administered with concentrations of the drugs that would otherwise far exceed toxicity thresholds if dosed continuously \cite{amin2010resiliency}. Continuous letrozole in combination with high dose intermittent ribociclib is currently in clinical trial (NCT02712723; ER-positive breast cancer) \cite{griffiths2021serial}. Intermittent high dose erlotinib delays resistance in an EGFR-mutant non–small cell lung cancer in vivo model \cite{chmielecki2011optimization,schottle2015intermittent}. Intermittent weekly EGFR-inhibitors reduced tumor load in vivo, compared with daily regimens with identical cumulative dose \cite{zhang2017effect}. Intermittent ``pulsatile'' high dose erlotinib once weekly maintains efficacy even after failure low dose continuous \cite{grommes2011pulsatile}. Ideal treatment protocols will maximize both first-order effects (cumulative dose) in addition to second-order effects (variance of dose delivered). Studies mentioned previously provide evidence of the tolerability of temporary dose escalation by also employing off-treatment periods to alleviate therapy toxicity.

\subsection{Defining (local) fragility in oncology}
Local Fragility, $F$, is a measurable quantity, similar to the Jensen Gap, defined as the difference in result of unevenness over evenness (with corresponding unevenness range parameter $\lambda$):
\begin{equation}
    F(x,\lambda) =  \frac{f(x+\lambda) + f(x-\lambda)}{2} - f(x) \label{fragility}
\end{equation}
What property of cancer cells, $f(x)$, is important in oncology? Here, we are interested in the gain of ``uneven'' high / low schedule over the ``even'' schedule. More precisely, fragility is the difference between 1) a schedule of two constant doses (termed an "even" dosing strategy), $\vec{X} = \{x, x\}$ and 2) a schedule of a high dose followed by a low dose (termed an "uneven" dosing strategy), $\vec{X} = \{x+\lambda, x-\lambda\}$. We consider the response to two doses over an interval of time $T$, where the first dose is given at $t=0$ and the second dose is given at $t=T/2$. Given a tumor with initial population size of $n_0$, the final size of an exponentially-growing population is given by:
\begin{eqnarray}
n_F(t) = n_0 \exp(\gamma(x) t),
\end{eqnarray}
\noindent where $\gamma(x)$ is the decay rate of the population associated with a dose of $x$. Fragility can be defined as:
\begin{equation}
    F(x,\lambda) =  n_0 \exp \left(\gamma(x+\lambda) \frac{T}{2}\right) \exp \left(\gamma(x-\lambda) \frac{T}{2} \right) - n_0 \exp \left(\gamma(x) \frac{T}{2}\right) \exp \left(\gamma(x) \frac{T}{2} \right),
\end{equation}
which simplifies to:
\begin{equation}
    F(x,\lambda) =  n_0 \left[ \exp \left( \left(\gamma(x+\lambda) + \gamma(x-\lambda) \right) \frac{T}{2} \right)  - \exp \left(\gamma(x) T \right)  \right].
\end{equation}
We are interested in the antifragile-fragile boundary, the point at which the population is no longer fragile but antifragile. If $F<0$, the cell population is antifragile by definition (a benefit conferred to uneven dosing for minimizing tumor growth rate) and if $F>0$, then fragile. It follows from the previous equation will be negative if:
\begin{equation}
    \frac{\gamma(x+\lambda) + \gamma(x-\lambda)}{2} < \gamma(x) \label{gamma}
\end{equation}
\color{black}Thus, the important domain for convexity is the dose-dependent growth rate, $\gamma(x)$. It is an unfortunate common practice to normalize dose response curves to obtain fractional survival at the final time point (e.g. $1-n_F/n_0$), which obscures convexity. Drug-induced growth rate inhibition (GR curves) have been introduced as a method to remove the artifactual dependency of IC50 and Emax on cellular division rate\cite{hafner2016growth}. GR-curves preserve convexity, unlike fractional survival.
\subsection{Fragility and Taylor Series Approximations}
Antifragility is a ``second-order'' effect. This is shown by first taking a Taylor Expansion about $x$:
\begin{equation}
    f(x \pm \lambda) = f(x) \pm \lambda f'(x) + \frac{\lambda^2}{2}f''(x) + O(\lambda^3)
\end{equation}
\noindent where $O(\lambda^3)$ represents all third-order or higher terms. Using this expansion, fragility can be written:
\begin{equation}
       F(x,\lambda) = \frac{1}{2} \bigg( 2f(x) - f(x) - \lambda f'(x) - \frac{\lambda^2}{2}f''(x) - f(x) + \lambda f'(x) - \frac{\lambda^2}{2}f''(x) \bigg) + O(\lambda^3).
\end{equation}
The zeroth-order terms and the first-order terms cancel out:
\begin{equation}
    F(x,\lambda) = -\frac{\lambda^2}{2}f''(x)+ O(\lambda^3)
\end{equation}
For small values of $\lambda$, fragility is proportional to the second derivative, and thus known as a second-order effect. Next, we connect the concept to finite difference methods.

\subsection{Fragility and finite differences}
\begin{figure}[t!]
\centering
\includegraphics[width=1\linewidth]{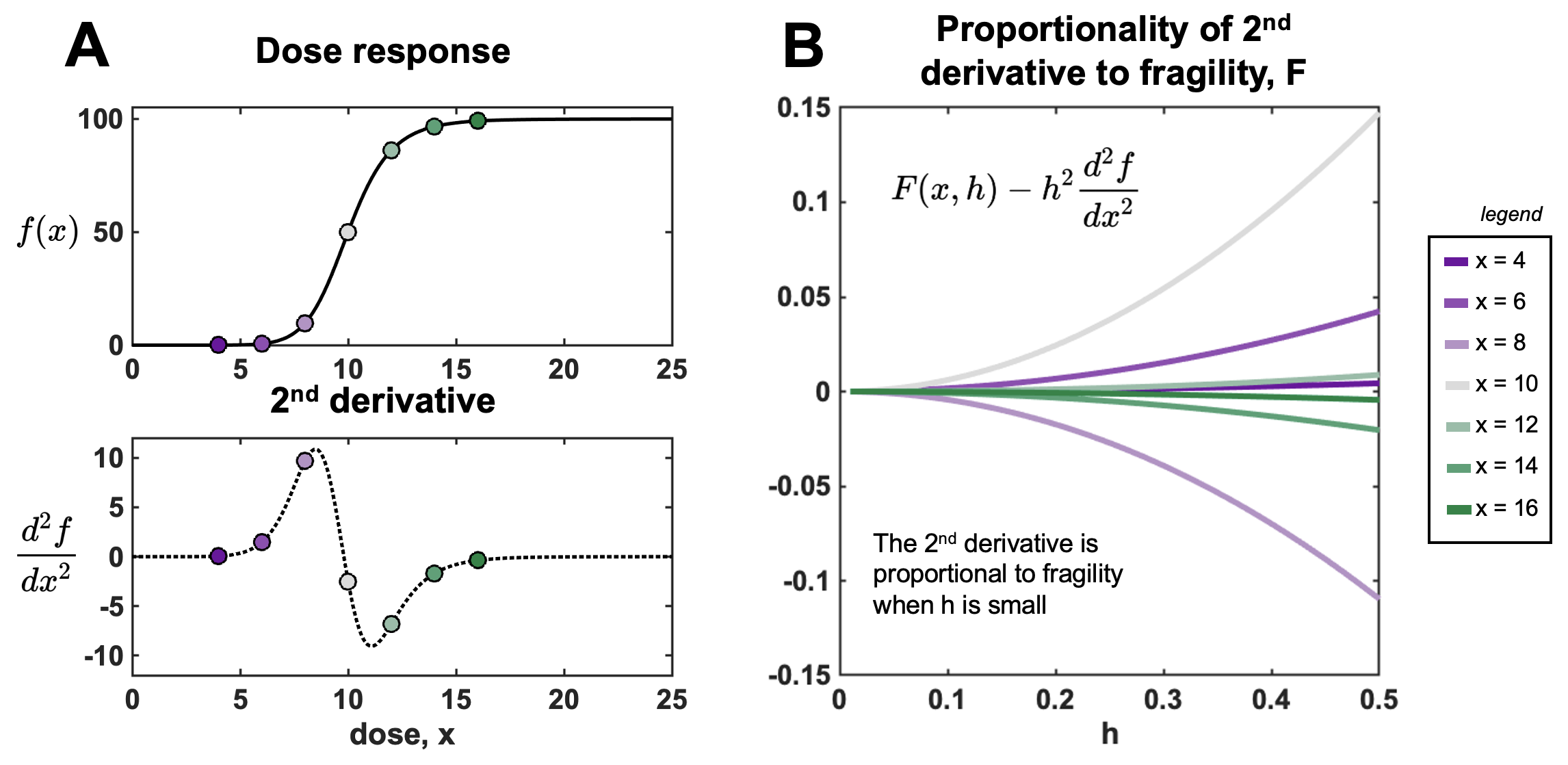}
\caption{ {\bf The second-derivative is an approximation for fragility for low values of h}. (A) Hill function, $H(x)$ (eqn. \ref{linear_Hill}) shown for $n=10$, $E_0=0$, $E_1=100$, $C=10$. Analytically derived second-derivative (eqn. \ref{ddH1}) is shown in the bottom panel. (B) Difference between fragility and second-derivative at various dose values (red to blue) corresponding to panel A. As $h\to0$, the error approaches zero: $F(x,h)-h^2 \frac{d^2H}{dx^2} \to 0$. }
\label{supp_second}
\end{figure}

We can approximate the derivative of $f(x)$ using finite differences. Here, we use the well-known central difference approximation to the second-derivative, $f''(x)$, where $h$ is the width of the interval over which the finite difference is estimated ($h>0$):\color{black}
% Here, we use central finite differences to obtain:
% \begin{eqnarray}
%     \frac{\Delta f}{\Delta x}=\frac{f\left(x+\frac{h}{2}\right)-f\left(x-\frac{h}{2}\right)}{h}\label{first_difference_quotient}.
% \end{eqnarray}
% This difference quotient is a first-order approximation of $f'(x)$ where $h$ is the width of the interval over which the finite difference is estimated ($h>0$). The approximation error is proportional to $h$. Taking the limit as $h\to0$ allows us to express $f'(x)$ in terms of this approximation:
% \begin{eqnarray}
%     f'(x)= \lim_{h\to0} \frac{f\left(x+\frac{h}{2}\right)-f\left(x-\frac{h}{2}\right)}{h}.
% \end{eqnarray}
% By iterating the central finite differences process once more, we can derive the finite difference quotient for $f'(x)$:
% \begin{eqnarray}
%     \frac{\Delta f'}{\Delta x}=\frac{f(x+h)-2f(x)+f(x-h)}{h^2}. \label{second_difference_quotient}
% \end{eqnarray}
% This difference quotient provides an approximation $f''(x)$ with an error proportional to $h^2$. Taking the limit as $h\to0$ yields the second derivative of $f(x)$:
\begin{eqnarray}
    f''(x)=\lim_{h\to0} \quad \frac{1}{h^2} \big [ f(x+h)+f(x-h)-2f(x) \big ].
\end{eqnarray}
The term on the right-hand side in brackets is related to fragility (eqn. \ref{fragility}), giving us the relationship between $F(x,\lambda=h)$ and $f''(x)$:
\begin{eqnarray}
    f''(x)=-\lim_{h\to0} \quad \frac{2}{h^2} F(x,h), \label{comparison}
\end{eqnarray}
This limit illustrates that $f''(x)$ approaches $F$ only when $h$ is small. When the value of $h$ is large, the approximation is poor, and therefore we employ equation \ref{fragility}. Figure \ref{supp_second} provides an example of approximation error for the Hill function (see next section).

\color{black}

\subsection{Applications to Hill function}\label{boundary_derivation}
The Hill function is commonly used to describe the drug pharmacodynamics\cite{meyer2019quantifying}, where $H(x)$ is the cell viability in response to a dose $x$. 
\begin{equation}
    H(x) = \frac{E_1-E_0}{1+\left(\frac{C}{x}\right)^n} + E_0 \label{linear_Hill}
\end{equation}
where $n$ is the Hill shape parameter, $E_0$ and $E_1$ are the minimal and maximal response (respectively), and $C$ is the half-maximal response (the EC50 value). The second derivative can be written:
\begin{equation}
    \frac{d^2H}{dx^2} = \frac{ (E_1 - E_0)nC^nx^{n-2} \big( (n-1)C^n-(n+1)x^n\big)}{ (C^n +x^n)^3}\label{ddH1}.
\end{equation}
Figure \ref{supp_second}A shows a sample Hill function and corresponding second derivative. Figure \ref{supp_second}B illustrates decreasing error between the numerical ($F$) and analytical ($\frac{d^2H}{dx^2}$) as $h\to0$. The error scales like $h^2$, as predicted in equation \ref{comparison}. The inflection point, found where $\frac{d^2H}{dx^2}=0$, defines the boundary between convex and concave regions.
\begin{equation}
    x^*=C \left ( \frac{n-1}{n+1} \right)^{1/n} \label{inflection}
\end{equation}
It can be shown that as $n$ increases, $x^*\rightarrow C$. Importantly, $x^*$ determines the boundary between antifragile and fragile regions of $f(x)$. Benefit can thus be derived from uneven dosing if $x<x^*$. The converse implies that uneven treatment schedules provide no additional benefit.  For a discussion on relaxing the assumption of fixed treatment schedules (e.g. figure \ref{Jensen}) using a probability density function describing dose distribution, see Appendix, equations \ref{a1} - \ref{a2}.
\color{black}

As shown in figure \ref{DoseResponseDist}, the input distribution passing through convex dose response (e.g. on the Hill function below its inflection point) result in a left-tailed outcome distribution, and a concave response function results in a right-tailed distribution.

\begin{figure*}
\includegraphics[width=\linewidth]{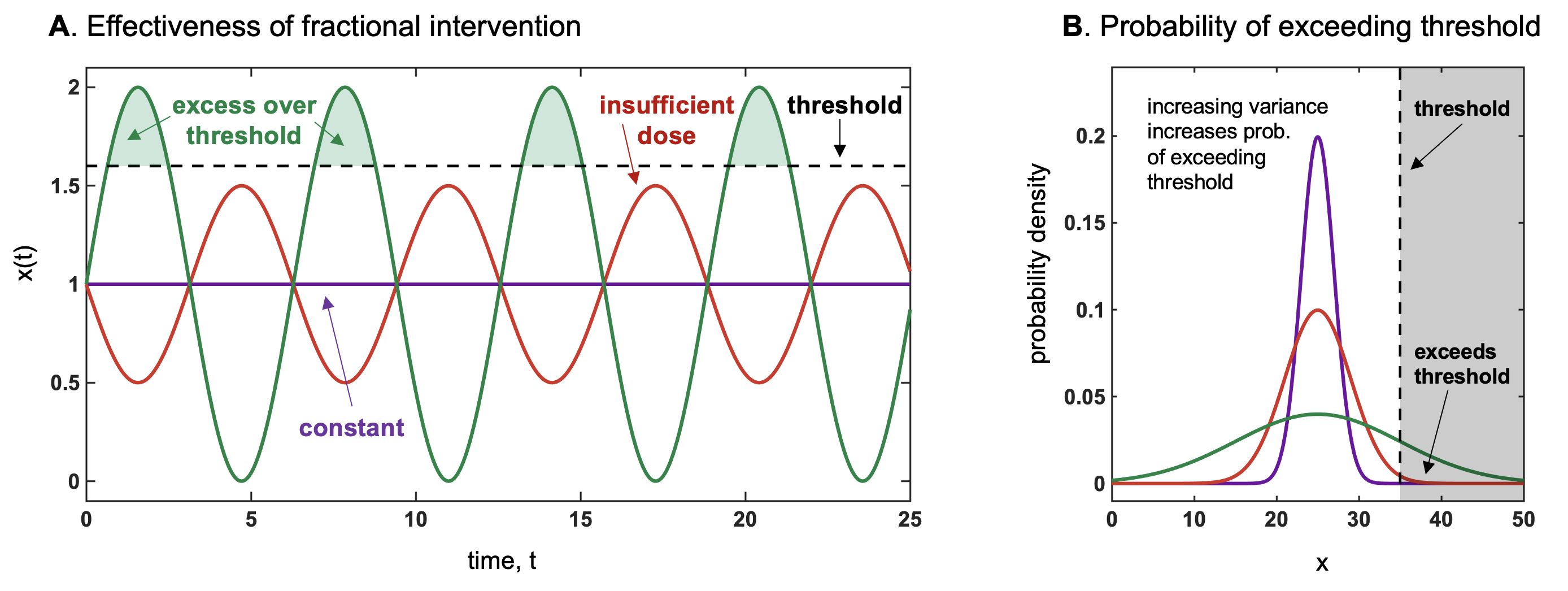}	
\caption{(A) How a fractional intervention is more effective to surpass a threshold than a constant dosage of the same average. This is akin to \textbf{stochastic resonance} (in physics) by which the presence of noise cause the signal to rise above the detection threshold. For instance, genetically modified BT crops produce a constant level of pesticide, which appears to be much less effective than occasional manual interventions to add doses to conventional plants. The same may apply to antibiotics, chemotherapy and radiation therapy\cite{cunningham2019call}. (B) How more variance impacts the exceedance over the threshold. If threshold $\geq$ mean, we have convexity and the variance increases the payoff more than variations in the mean. Such an effect is proportional to the remoteness of such threshold. Note that the harm function is defined as positive.}\label{dr36}
\end{figure*}
\begin{figure*} 
\includegraphics[width=\linewidth]{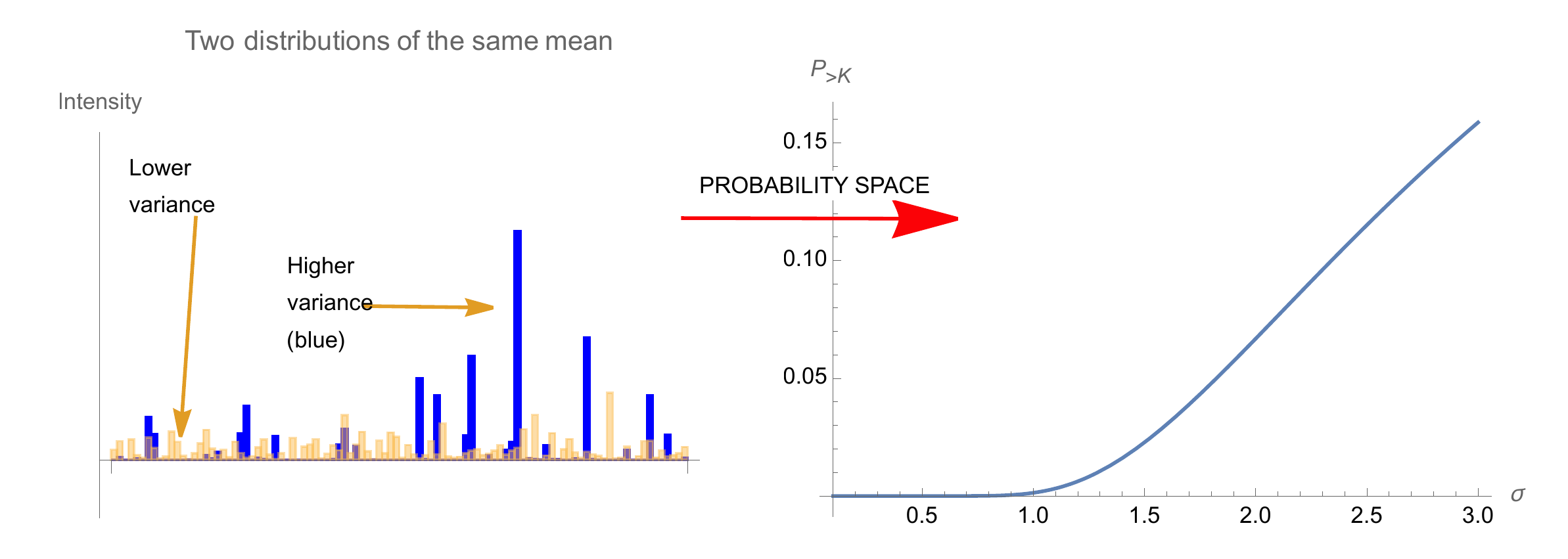}
\caption{(Left) A time series illustration of how a higher variance (hence scale), given the same mean, allow more spikes --hence an antifragile effect. We have random paths of two gamma distribution of same mean, different variances, $X_1\sim G(1,1)$ and $X_2\sim G(\frac{1}{10},10)$, showing higher spikes and maxima for $X_2$. The effect depends on norm $||.||_\infty $ , more sensitive to tail events, even more than just the scale which is  related to the norm $||.||_2 $.  (Right) Representation of Antifragility of (Left) in distribution space: we show the probability of exceeding a certain threshold for a variable, as a function of $\sigma$ the scale of the distribution, while keeping the mean constant. }\label{dr45}
\end{figure*}

%%%END REWRITE
Thus, we have defined as locally antifragile a situation in which, over a specific interval $[a,b]$,  either the expectation rises with the scale parameter of the probability distribution as in Eq. \ref{antifr}, or the dose response is convex (on average) over the same range. The designation in Taleb (2012) \cite{taleb2012antifragile} meant to accurately describe such situations: anything that gains from an increase in stochasticity or variability (since the scale parameter represents both). Terms like "resilience", since they were not mapped mathematically, are vague and even confusing as they meant either resistance or gains from stressors, depending on context.
Fig. \ref{dr36} and \ref{dr45} illustrates the threshold effect of the asymmetric response, and gives the intuition of how they can be described as as antifragile.

% Instead, if we restrict ourselves to the fixed treatment protocols found in figure \ref{Jensen}, we can define fragility, $F$, defined as the gain of ``uneven'' high / low schedule (with corresponding unevenness parameter $\sigma$) over the ``even'' schedule where $\sigma=0$. Expressed in terms of the hill function above, $F$ is:
% \begin{equation}
%     F(\bar{x},\sigma)= \frac{ H(\bar{x}+\sigma) + H(\bar{x}-\sigma)}{2} - H(\bar{x}) \label{fragility}
% \end{equation}
% If $F>0$, the dose response is antifragile by definition (benefit from unevenness), and if $F<0$, then fragile. Next, we turn to general sigmoid curves.

\begin{figure} 
	\includegraphics[width=\linewidth]{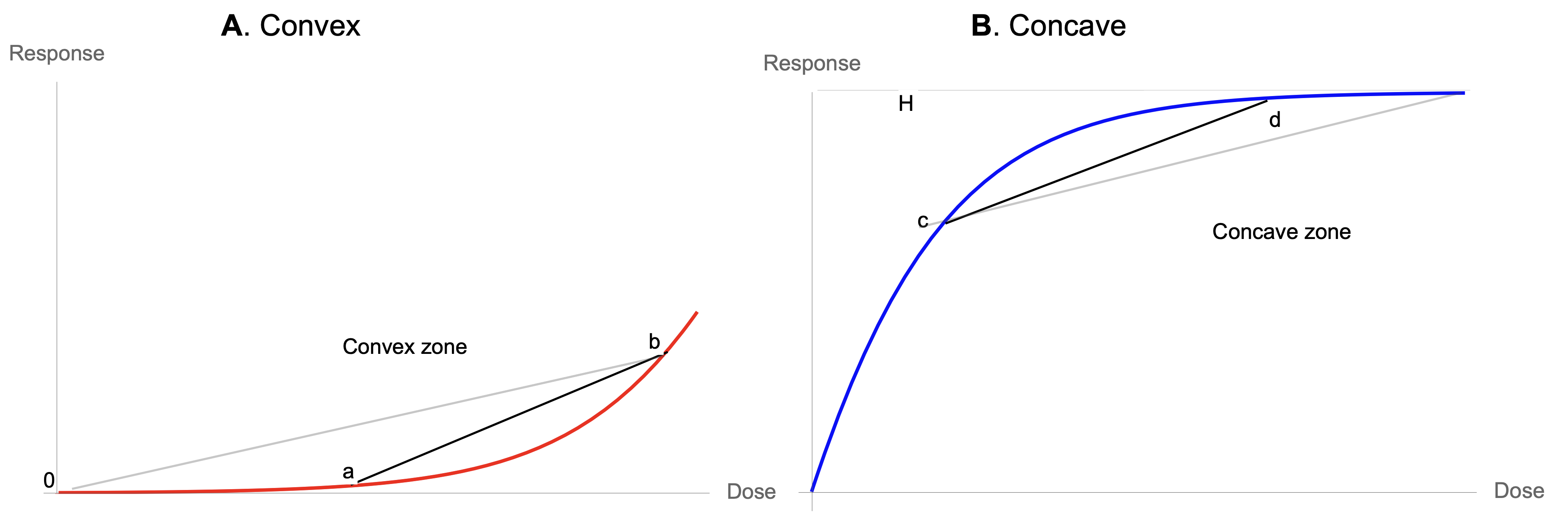}
	\caption{(A) Every (relatively) smooth dose-response with a floor has to be initially convex, hence prefers variations and concentration. (B) Every (relatively) smooth dose-response with a ceiling has to be concave while approaching the ceiling, hence prefers stability.} \label{dr12}
	\begin{center}
		\includegraphics[width=0.65\linewidth]{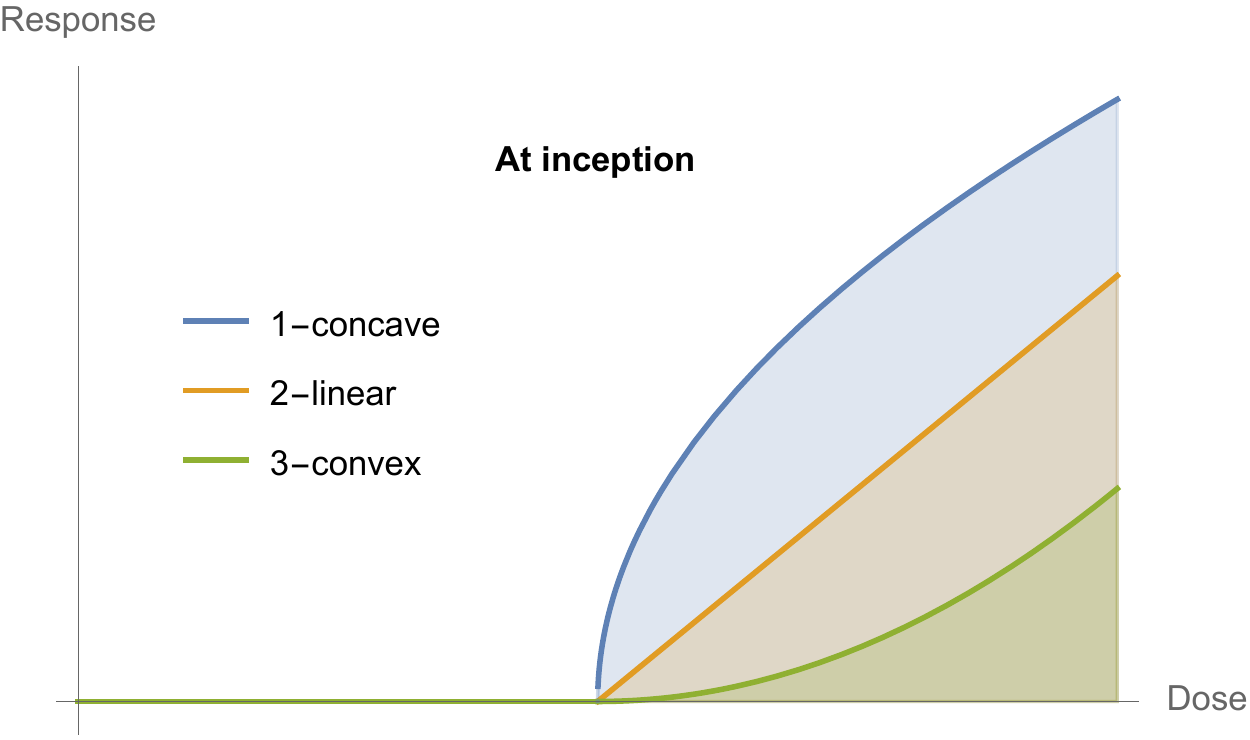}
	\caption{The three possibilities at inception} \label{threecurves}
	\end{center}
	\end{figure}
	
\subsection{ The first order sigmoid curve}
Next we outline the variety of sigmoids as catalogued by \cite{taleb2018anti}. Define the sigmoid or sigmoidal function as having membership in a class of function $\mathfrak{S}$, $S: \mathbb{R}\rightarrow [L,H]$, with additional membership in the $ \mathbb{C}^2$ class (twice differentiable), monotonic nonincreasing or nondecreasing, that is let $S'(x)$ be the first derivative with respect to $x$:  $S'(x) \geq 0$  for all $x$ or $S'(x) \leq 0$. We have:
$$ S(x) = \left\{ \begin{array}{ll}
         H & \mbox{as $x \rightarrow +\infty$};\\
        L & \mbox{if $x \rightarrow -\infty$}.\end{array} \right., $$
        which can of course be normalized with $H=1$ and $L=0$ if $S$ is increasing, or vice versa, or alternatively $H=0$ and $L=-1$ if $S$ is increasing. We can define the simple (or first order) sigmoid curve as having equal convexity in one portion and concavity in another: $\exists k >0 \text{ s.t. } \forall x_1<k \text{ and } x_2>k,$ $\text{sgn} \left(S''(x_1)\right) =-\text{sgn}(S''(x_2))$ if   $|S''(x_2)|\geq 0$.

Now all functions starting at 0 will have three possible properties at inception, as in Fig. \ref{threecurves}: concave, linear, convex. The point of our discussion is the latter becomes sigmoid and is of interest to us. Although few medical examples appear, under scrutiny, to belong to the first two cases, one cannot exclude them from analysis. We note that given that the inception of these curves is 0, no linear combination can be initially convex unless the curve is convex, which would not be the case if the start of the reaction is at level different from 0.

\begin{figure*}
\begin{center}
\includegraphics[width=.7\linewidth]{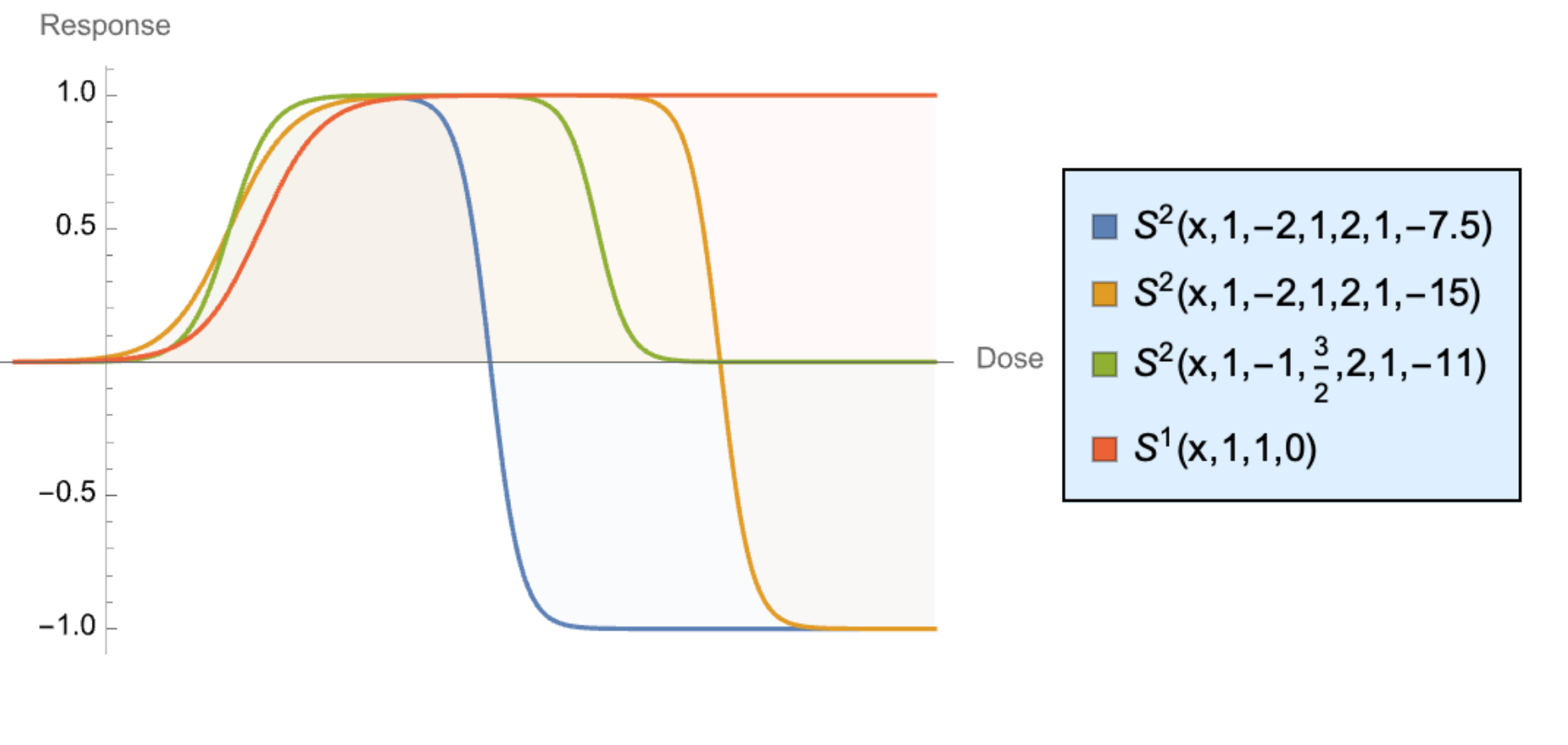}
\caption{Generalizing the Dose Response Curve, $S^2\left(x;a_1,a_2,b_1,b_2,c_1,c_2\right)\text{, }S^1\left(x;a_1,b_1,c_1\right)$ The convex part in the increasing section is what we call "antifragile". }\label{GDRCgraph}
\end{center}
\end{figure*}

There are many sub-classes of functions producing a sigmoidal effect. Examples: 
\begin{itemize}
	\item Pure sigmoids with smoothness characteristics expressed in trigonometric or exponential form, $f: \mathbb{R} \rightarrow [0,1]$: 
	$$f(x)=\frac{1}{2} \tanh \left(\frac{\kappa  x}{\pi }\right)+\frac{1}{2}$$ $$f(x)=\frac{1}{1-e^{-a x}}$$
	\item Gompertz functions (a vague classification that includes above curves but can also mean special functions )
	\item Special functions with support in $\mathbb{R}$ such as the Error function 
	$f: \mathbb{R} \rightarrow [0,1]$
	$$f(x)=-\frac{1}{2} \text{erfc}\left(-\frac{x}{\sqrt{2}}\right)$$
	
	\item Special functions with support in $[0,1]$, such as  $f: [0,1] \rightarrow [0,1]$	$$f(x)=I_x(a,b),$$
where $I_{(.)}(.,.) $ is the Beta regularized function.
    \item Special functions with support in $[0,\infty)$
   $$f(x)=Q\left(a,0,\frac{x}{b}\right)$$
   where $Q\left(.,.,.\right) $ is the gamma regularized function.   
	\item Piecewise sigmoids, such as the CDF of the Student Distribution
	$$ f(x)=\begin{cases}
 \frac{1}{2} I_{\frac{\alpha }{x^2+\alpha }}\left(\frac{\alpha
   }{2},\frac{1}{2}\right) & x\leq 0 \\
 \frac{1}{2} \left(I_{\frac{x^2}{x^2+\alpha }}\left(\frac{1}{2},\frac{\alpha
   }{2}\right)+1\right) & x> 0
\end{cases}$$

\end{itemize}

We note that the "smoothing" of the step function, or Heaviside theta $\theta(.)$ produces a sigmoid (in a situation of a distribution  or convoluted with a test function with compact support), such as $\frac{1}{2} \tanh \left(\frac{\kappa  x}{\pi }\right)+\frac{1}{2}$, with $\kappa  \rightarrow \infty$, see Fig. \ref{heaviside}.

\subsection{ Some necessary relations leading to a sigmoid curve}

	Let $f_1(x): \mathbb{R}^+\rightarrow [0,H]$ , $H\geq 0$, of class $C^2$ be the first order dose-response function, satisfying $f_1(0)=0$, $f_1'(0)|=0$, $\lim_{x \to +\infty} f_1(x)=H$, monotonic nondecreasing, that is, $f_1'(x)\geq 0 \;\forall x \in \mathbb{R}^+$, with a continuous second derivative, and analytic in the vicinity of $0$. Then we conjecture that: 	
	
	\textbf{A}- There is exist a zone $[0,b]$ in which $f_1(x)$ is convex, that is $f_1''(x)\geq 0$, with the implication that $\forall a \leq b$ a policy of variation of dosage produces beneficial effects: $$\alpha f_1(a) +(1-\alpha) f_1(b)\geq f_1(\alpha a+ (1-\alpha) b), 0\leq \alpha\leq 1.$$
	(The acute outperforms the chronic).
	
	\textbf{B}- There is exist a zone $[c,H]$ in which $f_1(x)$ is concave, that is $f_1''(x)\leq 0$, with the implication that $\exists d \geq c$ a policy of stability of dosage produces beneficial effects: $$\alpha f_1(c) +(1-\alpha) f_1(d)\leq f_1(\alpha c+ (1-\alpha) d).$$
	(The chronic outperforms the acute).
% 
	%\end{figure}

\section{The Generalized Dose Response Curve}\label{GDRC}
Let $S^N(x)$: $\mathbb{R}$ $\to $ [$k_L$, $k_R$], $S^N \in C^{\infty}$, be a continuous function possessing derivatives $\left(S^N\right)^{(n)}(x)$ of all orders, expressed as an $N$-summed and scaled standard sigmoid  functions:

\begin{equation}
S^N(x) \triangleq  \sum _{i=1}^N \frac{a_k}{1+e^{\left(-b_k x+c_k\right)}}\label{gensig}
\end{equation}
\noindent where $a_k, b_k, c_k$ are scaling constants $\in $ $\mathbb{R}$, satisfying: 
\begin{enumerate}
    \item $S^N$(-$\infty $) =$k_L$, and
    \item $S^N$($+\infty $) =$k_R$, and (equivalently for the first and last of the following conditions)
    \item $\frac{\partial ^2 S^N}{\partial x^2}$$\geq $ 0 { }for $x$ $\in $ (-$\infty $, $k_1$) , $\frac{\partial ^2 S^N}{\partial x^2}$$<$ 0 for $x$ $\in $ ($k_2$, $k_{>2}$), and $\frac{\partial ^2 S^N}{\partial x^2}$$\geq $ 0 for $x$ $\in $ ($k_{>2}$, $\infty $), 
with $k_1>k_2\geq \ldots \geq k_N$.
\end{enumerate}
By increasing $N$, we can approximate a continuous functions dense in a metric space, see Cybenko (1989) \cite{cybenko1989approximation}.

The shapes at different calibrations are shown in Fig. \ref{GDRCgraph}, in which we combined different values of N=2 $S^2\left(x;a_1,a_2,b_1,b_2,c_1,c_2\right)\text{, }$and the standard sigmoid  $S^1\left(x;a_1,b_1,c_1\right)$, with $a_1$=1, $b_1$=1 and $c_1$=0. As we can see, unlike the common sigmoid , the asymptotic response can be lower than the maximum, as our curves are not monotonically increasing. The sigmoid shows benefits increasing rapidly (the convex phase), then increasing at a slower and slower rate until saturation. Our more general case starts by increasing, but the reponse can be actually negative beyond the saturation phase, though in a convex manner. Harm slows down and becomes "flat" when something is totally broken.

\subsection{Antifragility and heterogeneity}
Tumors are composed of a heterogeneous collection of subpopulations with varied treatment sensitivity. Given $N$ non-interacting populations, the fragility of the total population is given by the sum of each subpopulation $i$'s fragility, $F_i$, weighted by it's frequency within the total population, $w_i$, such that $\sum_i w_i = 1$.
\begin{eqnarray}
    F(\bar{x},\sigma)&=& \sum_i^N w_i F_i(\bar{x},\sigma) \label{weighted_dose_response}
\end{eqnarray} 
\noindent where fragility $F_i$ for a single population is given by eqn. \ref{fragility}.

\begin{figure*}[h]
\centering
\includegraphics[width=1\linewidth]{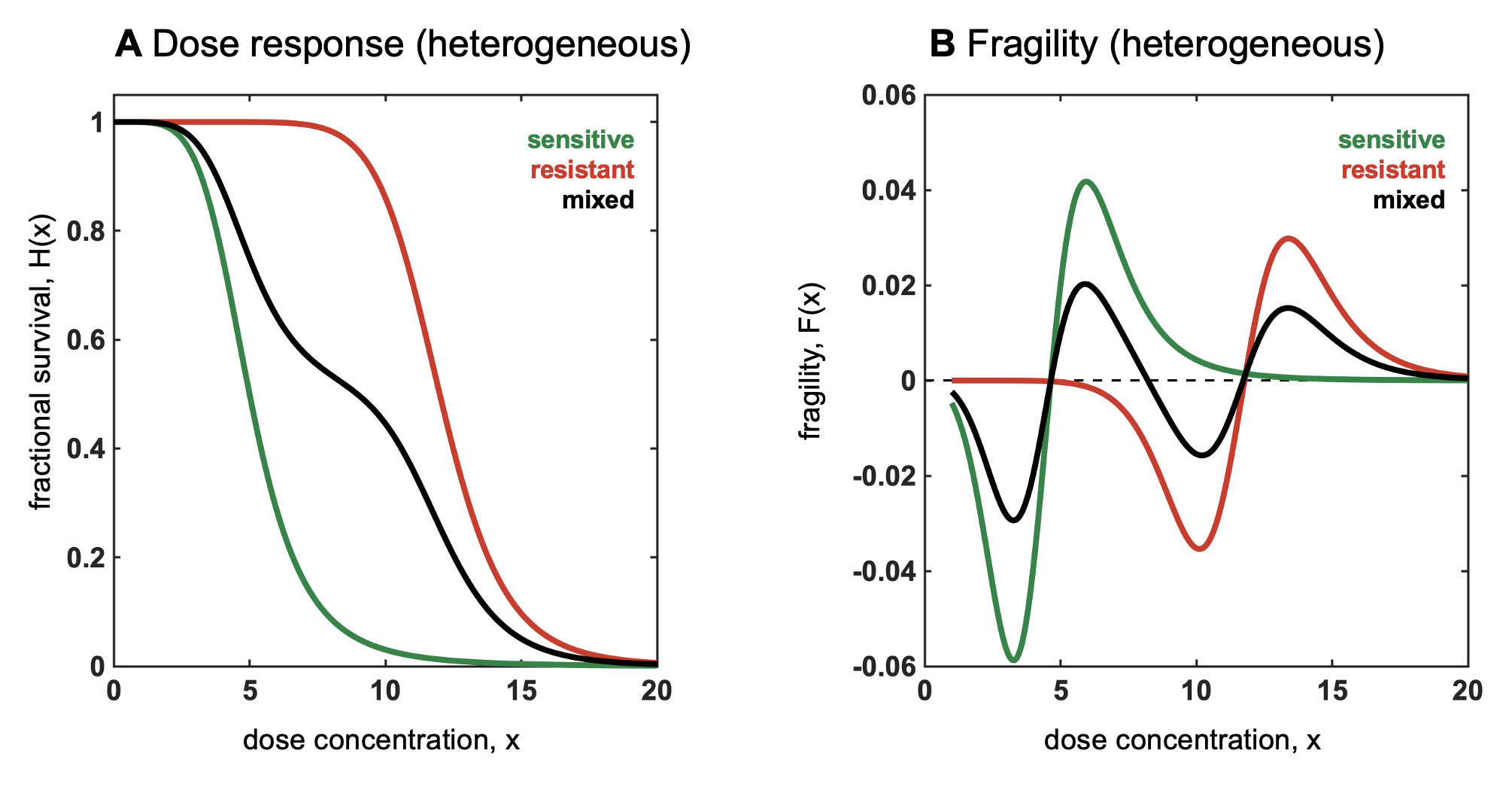}
\caption{ {\bf Relationship between convexity and mixed, heterogeneous populations} (A) Dose response shown for sensitive (green) and resistant (red) cell lines. When mixed, dose response is a weighted average of each (eqn. \ref{weighted_dose_response}; black)  (B) Fragility shown for sensitive (green) and resistant (red) cell lines. When mixed, fragility (black) switches from locally convex to locally concave multiple times. }
\label{resistance}
\end{figure*}

The simplest case of a heterogeneous mixture of two populations: sensitive (with associated dose response $H_1(x)$) and resistant ($H_2(x)$) is shown in Fig. \ref{resistance}A. In the case that each dose response, $H_i$, is non-increasing then the mixed dose response will also be non-increasing, but changes in convexity may occur. As seen in Fig. \ref{resistance}A, the mixed dose response has an internal plateau (shown for $w=0.5$). Local convexity (fragility) may switch signs multiple times.

\section{ Nonlinearities and Medical Iatrogenics}\label{iatrosection}
%\begin{figure} 
%	\includegraphics[width=0.65\linewidth]{Figures/talebdouadygr1.pdf}
%\caption{ A definition of fragility as left tail payoff sensitivity; the figure shows the effect of the perturbation of the lower semi-deviation $s^-$ on the tail integral ${\xi}$ of
%$(x -\Omega)$ below $K$, $\Omega$ being a centering constant. Our detection of fragility does not require the
%specification of $p$ the probability distribution.}\label{tailvegagraph}
%\end{figure}

Next we connect nonlinearity to iatrogenics (that is, harm done by the healer) for medicine in general, broadly defined as all manner of net deficit of benefits minus harm from a given intervention. 

The Taleb and Douady (2013) \cite{taleb2013mathematical} theorems state: 
\begin{itemize}
	
\item Convexity for a dose-response function increases fragility (from the expansion of the left tail in response to the increase in the scale of the distribution).	

\item Detection of a nonlinearity allows the prediction of fragility and helps formulate probabilistic decisions without much knowledge of the probability distribution --beyond minimum standard attributes.

\item The presence of concavity in the tails of the distribution implies a silent risk.

\end{itemize}
This approach was used in stress testing by the International Monetary Fund (IMF) where the degree of concavity in the tail was used as an indicator of the severity of tail exposure, see Taleb, Canetti et al. \cite{taleb2018IMF}.\footnote{ Such a method can transfer to medicine as the convexity of the  dose-response can be estimated via titration applied to Eq. \ref{fragility}.} %Simply compare initial dose plus $x$ to initial dose plus $2 x$ and check the presence of acceleration.}

\begin{figure} 
\includegraphics[width=\linewidth]{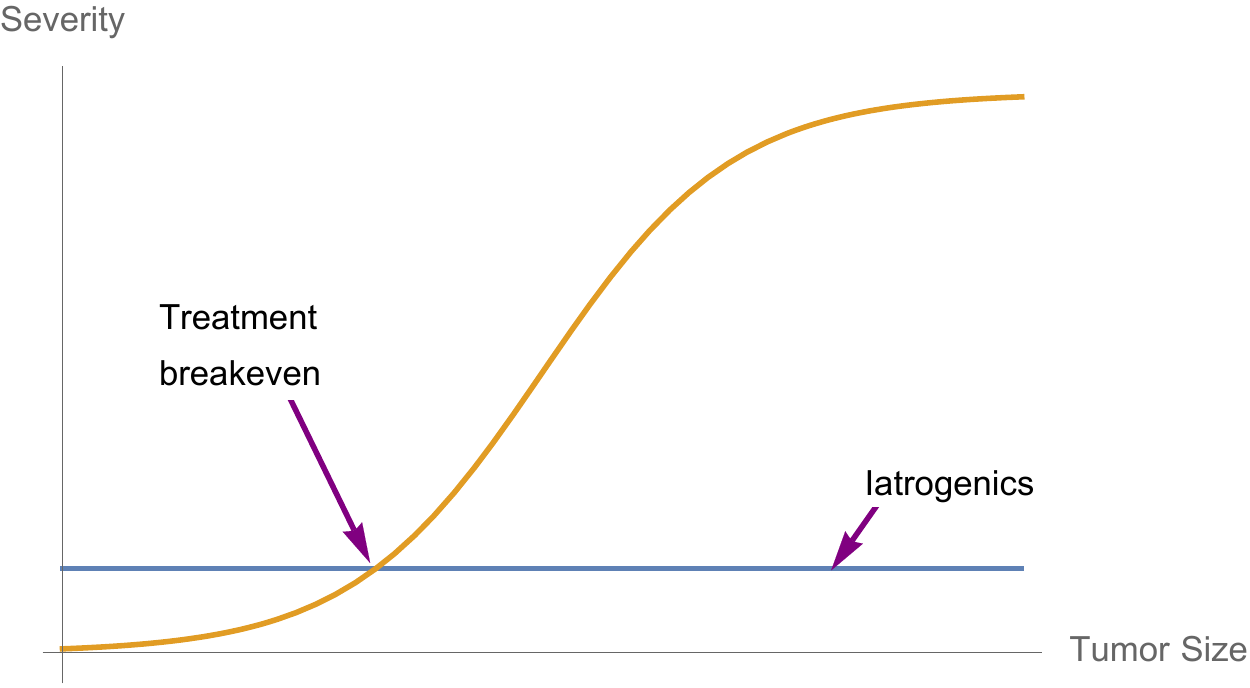}
\caption{Drug benefits when convex to Numbers Needed to Treat (NNT) in the left part, with gross iatrogenics invariant to condition (the constant line). We are assuming a standard sigmoidal benefit function.}\label{nnt}
\end{figure}

\begin{figure} 
\includegraphics[width=\linewidth]{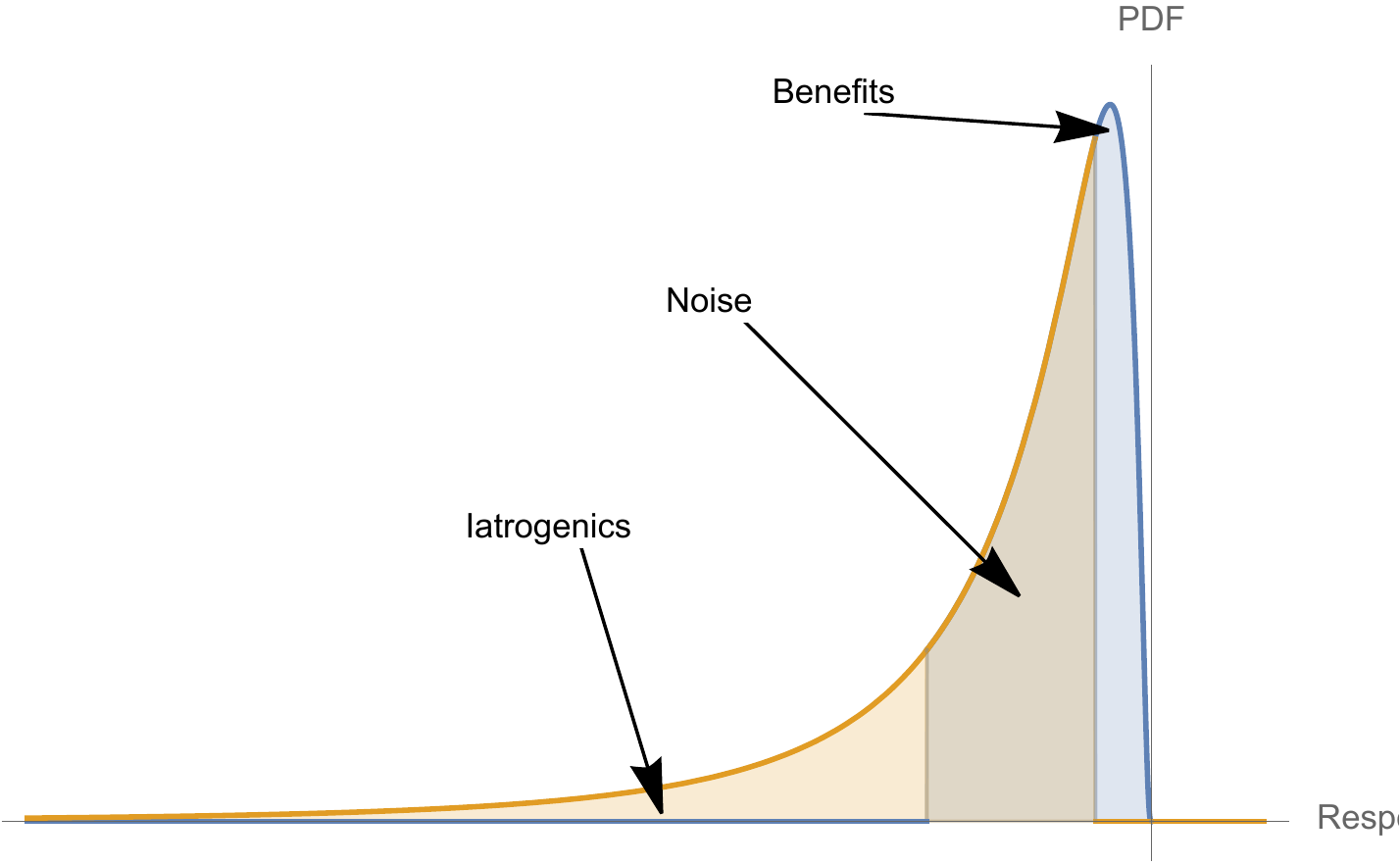}
\caption{Unseen risks and mild gains: translation of Fig. \ref{nnt} into a probabilistic representation, showing to the skewness of a decision involving iatrogenics when the condition is mild. This also gives the intuition of the Taleb and Douady\cite{taleb2013mathematical} translation theorems from concavity for $S(x)$ into a probabilistic attributes.}\label{skewness}	
\end{figure}

%\begin{figure}[h!]
%\includegraphics[width=0.65\linewidth]{Figures/concavityofhealthspending.jpg}
%\caption{Concavity of Gains to Health Spending. A more appropriate regression line than the one used by OECD should flatten off to the right, even invert to  fit the USA.  Credit: Edward Tufte}\label{healthstats}
%\end{figure}

\subsection{ Effect reversal}
Radiation might be beneficial in small doses, with reversal later on. In Neumaier et al. (2012) \cite{neumaier2012evidence} titled "Evidence for formation of DNA repair centers and dose-response nonlinearity in human cells":
\begin{quotation}
	The standard model currently in use applies a linear scale, extrapolating cancer risk from high doses to low doses of ionizing radiation. However, our discovery of DSB clustering over such large distances casts considerable doubts on the general assumption that risk to ionizing radiation is proportional to dose, and instead provides a mechanism that could more accurately address risk dose dependency of ionizing radiation.
\end{quotation}
 So low-level radiation may cause hormetic overreaction producing protective effects. Also see Tubiana et al. (2005) \cite{tubiana2006recent}. Bharadwaj and Stafford (2010) present similar general-sigmoidal effects in hormonal disruptions by chemicals \cite{bharadwaj2010hormones}.

\subsection{ Nonlinearity of NNT and the consequences}
Below are applications of convexity analysis in decision-making in dosage, shown in Fig. \ref{nnt} and Fig. \ref{skewness}.  In short, it is fallacious to translate a policy derived from acute conditions and apply it to milder ones. Mild conditions are different in treatment from an acute one. Likewise, high risk is qualitatively different from mild risk.
 
There is an active literature on "overdiagnosis", see Kalager et al(2012) \cite{kalager2012overdiagnosis}, Morell et al.(2012) \cite{morrell2010estimates}. The point is that treating a tumor that doesn't kill reduces life expectancy; hence the need to balance iatrogenics and risk of cancer. An application of nonlinearity can shed some light to the approach and clarify the public debate \cite{taleb2012antifragile}.

In a similar spirit of avoiding over-treatment, adaptive therapy in metastatic castrate-resistant prostate cancer (clinical trial NCT02415621) has illustrated the feasibility of irregular treatment protocols based on algorithms that react to tumor response. Adaptive treatment protocols maintain a stable population of sensitive cells in order to suppress the emergence of resistance \cite{gatenby2009adaptive, west2020towards}. Resistance in some cases is similar to the irreconciliable ruin of a financial ``blowup'' as patients may develop multi-drug resistance to structurally or functionally different drugs. The irreversibility of such clinical outcomes is similar to that of financial ruin, a \textit{tail fragility} situation from which the agent cannot exit. Adaptive algorithms decrease the cumulative dose administered to a patient, lessening the selection for resistance \cite{cunningham2018optimal}. While it is not an explicitly stated goal of adaptive therapy, these schedules increase both intra- and inter-patient dosing variance \cite{zhang2017integrating}.
\footnote{
This year (2022) saw the publication of calls from within the FDA to revamp the dose-finding protocols to be suitable for targeted therapies\cite{cancer_letter_fda}. Traditional dose selection protocols invented for use with cytotoxic chemotherapies may not apply to targeted therapies that have exposure-response curves which plateau at low toxicities to the patient. Differences in convexity between chemotherapies and newly developed targeted therapies leading to differing outcomes in diminished returns of dose escalation and differing curvature of dose response curves.
}

%\bibliographystyle{IEEEtran}
%\bibliography{/Users/nntaleb/Dropbox/Central-bibliography}

\vspace{6pt} 

\appendix

\section[\appendixname~\thesection]{Antifragility indirectly detected in the various literatures}

%Let us review the various literature that found benefits in increase in scale (i.e. local antifragility) though without gluing their results as part of a general function.% In short the papers in this section show \textit{indirectly} the effects of an increase in $\sigma$ for diabetes, alzheimer, cancer rates, or whatever condition they studied. The scale of the distribution means increasing the variance, say instead of giving a feeding of $x$ over each time step $\Delta t$, giving $x -\delta$ then $x +\delta$ instead, as in Eqs. \ref{jensen} and \ref{antifr}.  Simply, intermittent fasting would be having $\Delta \approx x$. and the scale can be written in such a simplified example as the dispersion $\sigma \approx \delta$. 
  
% 
%\begin{figure} [h!]
%	\includegraphics[width=0.65\linewidth]{Figures/kaiser2003hormesis.png}
%\caption{ Hormesis in Kaiser (2003) we can detext a convex-concave sigmoidal shape that fits our generalized sigmoid in Eq.\ref{gensig}.}\label{kaiser}
%\end{figure}

Table \ref{tab1} reviews the medical literature on embedded antifragility, defined as indirectly producing evidence of benefits from the "disorder cluster".
\begin{table}
\caption{Review of Medical Research on Embedded Antifragility\label{tab1}}
%	\begin{adjustwidth}{-\extralength}{0cm}
%		\newcolumntype{C}{>{\centering\arraybackslash}X}
	%	\begin{tabularx}{\fulllength}{CC}
%			\toprule
%\begin{tabular}{cc}
%			\textbf{Field}	& \textbf{Papers}	   \\
%			\midrule
			\textbf{Mithridatization and hormesis }		 Kaiser (2003) \cite{kaiser2003sipping} %(see Fig. \ref{kaiser})
			, Rattan (2008) \cite{rattan2008hormesis}, Calabrese and Baldwin (2002, 2003a, 2003b) \cite{calabrese2002defining},\cite{calabrese2003hormesis},\cite{calabrese2003hormetic}, Aruguman et al (2006) \cite{arumugam2006hormesis}.					\\
			
			\textbf{Caloric restriction and hormesis } 	 Martin, Mattson et al. (2006) \cite{martin2006caloric}				 \textsuperscript{1}\\
			
			\textbf{Cancer treatment and fasting }  Longo et al. (2010) \cite{longo2010calorie}, Safdie et al. (2009) \cite{safdie2009fasting}, Raffaghelo et al. (2010), \cite{raffaghello2010fasting}, Lee et al (2012) \cite{lee2012fasting}\\
			
{\bf Aging and intermittence }  Fontana et al. \cite{fontana2014medical} \\
\\
 {\bf For brain effects }  Anson, Guo, et al. (2003) \cite{anson2003intermittent}, Halagappa, Guo, et al. (2007) \cite{halagappa2007intermittent}, Stranahan and Mattson (2012) \cite{stranahan2012recruiting}. The long-held belief that the brain needed glucose, not ketones, and that the brain does not go through autophagy, has been progressively replaced\\
 \\

   {\bf On yeast and longevity under restriction } Fabrizio et al. (2001)\cite{fabrizio2001regulation}; SIRT1, Longo et al. (2006) \cite{longo2006sirtuins}, Michan et al. (2010) \cite{michan2010sirt1}\\
\\
    {\bf For diabetes, remission or reversal } Taylor (2008) \cite{taylor2008pathogenesis}, Lim et al. (2011) \cite{lim2011reversal}, Boucher et al. (2004) \cite{boucher2004biochemical}; diabetes management by diet alone, early insights in Wilson et al. (1980) \cite{wilson1980dietary}.  Couzin (2008) \cite{couzin2008deaths} gives insight that blood sugar stabilization does not have the effect anticipated (there needs to be stressors). The ACCORD study (Action to Control Cardiovascular Risk in Diabetes) found no benefits from lowering blood glucose levels. Synthesis, Skyler et al. (2009) \cite{skyler2009intensive}, old methods, Westman and Vernon (2008) \cite{westman2008has}. Bariatric (or other) surgery as alternative to intermittent fasting: Pories (1995) \cite{pories1995would}, Guidone et al. (2006) \cite{guidone2006mechanisms}, Rubino et al. 2006 \cite{rubino2006mechanism}\\
%    \\
%    %	\noindent{\footnotesize{\textsuperscript{1} This is a table footnote.}}
%\bottomrule
%		\end{tabularx}
%	\end{adjustwidth}
%%	\noindent{\footnotesize{\textsuperscript{1} This is a table footnote.}}
%\end{table}
%
%\begin{table}[H]
%\caption{Review of Medical Research on Intermittence (cont)\label{tab2}}
%	\begin{adjustwidth}{-\extralength}{0cm}
%		\newcolumntype{C}{>{\centering\arraybackslash}X}
%		\begin{tabularx}{\fulllength}{CCCC}
%			\toprule
%			\textbf{Field}	& \textbf{Papers}	  
%			\midrule
   \\      
     {\bf Ramadan and effect of fasting  } Trabelsi et al. (2012) \cite{trabelsi2012effect}, Akanji et al. (2012). Note that the Ramadan time window is short (12 to 17 hours) and possibly fraught with overeating so conclusions need to take into account energy balance and that the considered effect is at the low-frequency part of the timescale\\
     \\
    
   {\bf Caloric restriction}  Harrison (1984), Wiendruch (1996), Pischon (2008)\\
   \\
    
   {\bf Autophagy for cancer} Kondo et al. (2005) \cite{Kondo2005role}\\
   \\   
     {\bf Autophagy (general)} Danchin et al. (2011) \cite{danchin2011antifragility}, He et al. (2012) \cite{he2012exercise}\\
    \\
   {\bf Fractional dosage} Wu et al. (2016) \cite{wu2016fractional}\\
   \\   
  {\bf Jensen's inequality in exercise}  Many such as Schnohr and Marott (2011) \cite{schnohr2011intensity},intermittent extremes vs moderate physical activity\\
  \\
 {\bf Cluster of ailments}  Yaffe and Blackwell (2004) \cite{yaffe2004diabetes}, Alzheimer and hyperinsulenemia, Razay and Wilcock (1994) \cite{razay1994hyperinsulinaemia};  Luchsinger, Tang, et al. (2002) \cite{luchsinger2002caloric}, Luchsinger Tang et al. (2004) \cite{luchsinger2004hyperinsulinemia} Janson, Laedtke, et al. (2004) \cite{janson2004increased}.
 \\ 
    \\    
 {\bf Benefits  of \textit{some type of} stress (and convexity of the effect)} For the different results from the two types of stressors, short and chronic, Dhabar (2009) "A hassle a day may keep the pathogens away: the fight-or-flight stress response and the augmentation of immune function" \cite{dhabhar2009hassle}.  for the benefits of stress on boosting immunity and cancer resistance (squamous cell carcinoma), Dhabhar et al. (2010) \cite{dhabhar2010short}, Dhabhar et al. (2012) \cite{dhabhar2012high} , Ansbacher et al. (2013)\cite{aschbacher2013good}\\
 \\      
   {\bf Iatrogenics of hygiene and systematic elimination of germs} Rook (2011) \cite{rook2011hygiene}, Rook (2012) \cite{rook2012hygiene} (auto-immune diseases from absence of stressor), M\'egraud and Lamouliatte (1992) \cite{megraud1992helicobacter} for Helyobacter Pilori and incidence of cancer\\
%			\bottomrule
%\end{tabular}
%		\end{tabularx}
%	\end{adjustwidth}
%	\noindent{\footnotesize{\textsuperscript{1} This is a table footnote.}}
\end{table}
\section[\appendixname~\thesection]{Simple Convexity and its Effects}\label{convex1}

To eliminate ambiguity, let us define convexity. Let $f(.)$ be  the response function, $f:\mathbb{R}^+ \rightarrow \mathbb{R} $ be a twice differentiable function. If over a range x $\in [a,b]$, over a set time period $\Delta t$, $\frac{\partial  ^2f(x)}{\partial  x^2}\geq $ 0, or more practically (by relaxing the assumptions of differentiability), $\frac{1}{2}(f(x+\Delta x) + f(x-\Delta x))\geq f(x)$, with $x+\Delta x$ and $x-\Delta x \in [a,b]$ then there are benefits or harm from the unevenness of distribution, pending whether $f$ is defined as positive or favorable or modeled as a harm function (in which case one needs to reverse the sign for the interpretation).

We can generalize to comparing linear combinations:
$\sum\alpha_i =1$, $0\leq|\alpha_i|\leq 1$, $\sum(\alpha_i f(x_i)) \geq f(\sum(\alpha_i x_i))$; thus we end up with situations where, for $ x \leq b-\Delta$ and $n \in \mathbb{N}$, $f(n x) \geq n f(x)$. This last property describes a "stressor" as having higher intensity than zero: there may be no harm from $f(x)$ yet there will be one at higher levels of $x$.

Now if $X$ is a random variable with support in $[a,b]$ and $f$ is convex over the interval as per above, then 
\begin{equation}
 \mathbb{E}\left(f(x)\right) \geq f\left(\mathbb{E}(x)\right)\label{jensen}	, 
 \end{equation}
 what is commonly known as Jensen's Inequality, see Jensen(1906) \cite{jensen1906fonctions}, Fig. \ref{Jensen}. Further (without loss of generality), if its continuous distribution with density $\varphi(x)$ and support in $[a,b]$ belongs to the location scale family distribution, with $\varphi(\frac{x}{\sigma})= \sigma \varphi(x)$ and $\sigma>0$, then, with  $\mathbb{E}_{\sigma}$ the indexing representing the expectation under a probability distribution indexed by the scale $\sigma$, we have:
\begin{equation}
	\forall \sigma_2>\sigma_1,\, \mathbb{E}_{\sigma_2}\left(f(x)\right) \geq \mathbb{E}_{\sigma_1}\left(f(x)\right) \label{antifr}
\end{equation}
The last property implies that the convexity effect increases the expectation operator. We can verify that since $\int_{f(a)}^{f(b)} y \frac{\phi \left(f^{(-1)}(y)\right)}{f'\left(f^{(-1)}(y)\right)}\mathrm{d}y$
 is an increasing function of $\sigma$. A more simple approach (inspired from mathematical finance heuristics) is to consider for $0\leq\delta_1 \leq \delta_2\leq b-a$, where $\delta_1$ and $\delta_2$ are the mean expected deviations or, alternatively, the results of a simplified two-state system, each with probability $\frac{1}{2}$:
 \begin{equation}
 \frac{f(x-\delta_2)+	 f(x+\delta_2)}{2}\geq  \frac{f(x-\delta_1)+	 f(x+\delta_1)}{2} \geq f(x)
 \end{equation}

This is of course a simplification here since dose response is rarely monotone in its nonlinearity, as we will see in later sections. But we can at least make claims in a certain interval $[a,b]$ and it can produce useful heuristics.

What are we measuring? Clearly, the dose (represented on the $x$ line) is hardly ambiguous: any measurable quantity can do, such as systolic blood pressure, ejection fraction, caloric deficit, pounds per square inch, temperature, etc. The response, harm or benefits, $f(x)$ on the other hand, need to be equally precise, nothing vague, such as hazard ratios, some quantifiable index of health, median life expectancy, and similar quantities. If one cannot express the response quantitatively, then such an analysis cannot apply. 

\section[\appendixname~\thesection]{Relaxing the assumption of fixed treatment schedules}
We can relax the assumption of fixed treatment schedules (e.g. figure \ref{Jensen}). Given some input probability density function describing the distribution of dose, $p(x)$, the probability density can be determined for the Hill function analytically. Let $X$ (dose concentration) and $Y = H(X)$ be random variables. The probability density function transformation, $P(y(a) \leq Y < y(b) )$, is given by:
\begin{eqnarray}
    P &=& \int_a^b p(x) dx = \int_{y(a)}^{f(b)} p(x(y)) \left | \frac{dx}{dy} \right | dy \label{a1} \\
    &=& \int_a^b p \left ( C \left( \frac{E_1-E_0}{y-E_0} \right) ^{\frac{-1}{n}} \right )  \left | \frac{C(E_1-E_0)}{n(E_0-y)^2} \left ( \left( \frac{E_0-E_1}{E_0-y} \right ) - 1 \right ) ^{\frac{-(n+1)}{n}} \right | dy \label{a2}
\end{eqnarray}

\section{conflictsofinterest}
The authors declare no conflict of interest.

\section{acknowledgments}
Yaneer Bar Yam and participants in the International Conference On Complex Systems, Boston, 2018 where some early ideas leading to this paper were presented \cite{taleb2018anti} (synthesized in the appendices), as well as Raphael Douady, Luke Pierik, Sandy Anderson, Maximilian Strobl, and Andriy Marusyk. Editorial Note: excluding appendix and supplementary material, the overlap with this document and the conference proceedings is below 20\%.

%%%%%%%%%%%%%%%%%%%%%%%%%%%%%%%%%%%%%%%%%%
%\begin{adjustwidth}{-\extralength}{0cm}

%\end{adjustwidth}
\end{document}